\documentclass[twocolumn,superscriptaddress,prl]{revtex4-2}

\usepackage{graphicx}
\usepackage{color}
\usepackage{soul} 
\usepackage{amsmath}
\usepackage{hyperref}
\usepackage{braket}
\usepackage{mathtools}
\usepackage{lipsum}
\hypersetup{colorlinks=true, citecolor=cyan, urlcolor=blue, linkcolor=blue}
\usepackage[capitalize]{cleveref}
\DeclareUnicodeCharacter{2009}{\,} 

\usepackage{graphicx,subfigure,amsmath,bbm}
\usepackage[utf8]{inputenc}
\usepackage{mathtools}
\usepackage[makeroom]{cancel}

\newcommand{\new}[1]{\textcolor{blue}{#1}}

\newcommand{\RMP}[4]{\textit{#1}, Rev. Mod. Phys. \textbf{#2}, #3 (#4)}
\newcommand{\PR}[4]{\textit{#1}, Phys. Rev. \textbf{#2}, #3 (#4)}
\newcommand{\PRXQuantum}[4]{\textit{#1}, PRX Quantum \textbf{#2}, #3 (#4)}
\newcommand{\PRL}[4]{\textit{#1}, Phys. Rev. Lett. \textbf{#2}, #3 (#4)}
\newcommand{\PRR}[4]{\textit{#1}, Phys. Rev. Research. \textbf{#2}, #3 (#4)}
\newcommand{\PRA}[4]{\textit{#1}, Phys. Rev. A \textbf{#2}, #3 (#4)}
\newcommand{\PRB}[4]{\textit{#1}, Phys. Rev. B \textbf{#2}, #3 (#4)}
\newcommand{\PRApplied}[4]{\textit{#1}, Phys. Rev. Appl. \textbf{#2}, #3 (#4)}
\newcommand{\PRX}[4]{\textit{#1}, Phys. Rev. X \textbf{#2}, #3 (#4)}
\newcommand{\Science}[4]{\textit{#1}, Science \textbf{#2}, #3 (#4)}
\newcommand{\Nature}[4]{\textit{#1}, Nature \textbf{#2}, #3 (#4)}

\newcommand{\NComm}[4]{\textit{#1}, Nat. Comm \textbf{#2}, #3 (#4)}
\newcommand{\SciAdv}[4]{\textit{#1}, Sci. Adv \textbf{#2}, #3 (#4)}

\newcommand{\NatureNano}[4]{\textit{#1}, Nat. Nanotech \textbf{#2}, #3 (#4)}

\begin{document}

\title{Resolving non-perturbative renormalization of a microwave-dressed weakly anharmonic superconducting qubit coupled to a single quantized mode}

\author{Byoung-moo Ann}
\email{byoungmoo.ann@gmail.com}
\affiliation{Kavli Institute of Nanoscience, Delft University of Technology, 2628 CJ Delft, The Netherlands} 
\affiliation{Quantum Technology Institute, Korea Research Institute of Standards and Science, 34113 Daejeon, South Korea}
\author{Sercan Deve}
\affiliation{Kavli Institute of Nanoscience, Delft University of Technology, 2628 CJ Delft, The Netherlands} 
\author{Gary A. Steele}
\affiliation{Kavli Institute of Nanoscience, Delft University of Technology, 2628 CJ Delft, The Netherlands} 
\date{\today}

\begin{abstract}
%
Microwave driving is a ubiquitous technique for superconducting qubits (SCQs), but the dressed states description based on the conventionally used perturbation theory cannot fully capture the dynamics in the strong driving limit. Comprehensive studies beyond these approximations applicable to transmon-based circuit quantum electrodynamics (QED) systems are unfortunately rare as the relevant works have been mainly limited to single-mode or two-state systems.
In this work, we investigate a microwave-dressed transmon coupled to a single quantized mode over a wide range of driving parameters.
We reveal that the interaction between the transmon and resonator as well as the properties of each mode is significantly renormalized in the strong driving limit.
Unlike previous theoretical works, we establish a non-recursive, and non-Floquet theory beyond the perturbative regimes, which excellently quantifies the experiments.
This work expands our fundamental understanding of dressed cavity QED-like systems beyond the conventional approximations.
Our work will also contribute to fast quantum gate implementation, qubit parameter engineering, and fundamental studies on driven nonlinear systems.
%
%
\end{abstract}

\maketitle
Dynamically driving systems is a common methodology in physics \cite{driven-system-0, driven-system-1,driven-system-2,driven-system-3,driven-system-4,driven-system-5,driven-system-6,driven-system-7,driven-system-8}.
Qubits or oscillators driven by time-periodic potentials are the most typical type of such systems.
In circuit quantum electrodynamics (QED) platforms, a prototypical system for exploring and understanding light-matter interactions in the quantum regime \cite{cQED}, applying time-periodic potentials through charge or flux lines is a major means to perform quantum gate operations \cite{driven-system-9, driven-system-10, driven-system-11, driven-system-12, driven-system-13, driven-system-14, driven-system-15,Strand-PRB-2013, Krinner-PRApplied-2020}, or engineer the qubit's properties \textit{in-situ} \cite{QuEngineering-1,QuEngineering-2,QuEngineering-3,QuEngineering-4}.

In the strong driving limit, the significantly renormalized eigenbasis of multi-level qubits cannot be captured by low-order perturbation theory (PT), and two-state (TS) description.
These can modify the quantum dynamics quantitatively and qualitatively. Unfortunately, investigating circuit QED or even general cavity QED-like systems in this direction has remained unexplored, although periodically driven quantum systems in the strong drive limit have been explored in various platforms \cite{Dressed-tls-1,Dressed-tls-2,Dressed-tls-3,Dressed-tls-4,Dressed-tls-5,Dressed-tls-6,Dressed-tls-7,Dressed-tls-8,Dressed-tls-9,Dressed-tls-10,Dressed-tls-11,Dressed-tls-12,Dressed-tls-13,Dressed-tls-14,Dressed-tls-15,Dressed-tls-16,Dressed-tls-17,Dressed-tls-18}.

Here, we perform a study on the renormalization of a transmon coupled to a resonator.
In-depth investigations of the driven transmon–resonator configuration have been intensively performed \cite{multi-1,multi-3,multi-4,multi-5,multi-6,multi-6-1,multi-6-2,multi-6-3}.
Recently, the efforts to break conventional approximations are also being actively reported \cite{multi-2, multi-7, multi-8,multi-9,multi-10,multi-11,multi-12,multi-13,multi-14,multi-15}.
Unlike most of the previous studies, we derive a non-recursive and non-Floquet formula, advantageous for multi-level systems and non-perturbative problems.
We identify a non-perturbatively modified qubit–resonator interaction in the experiments through Lamb shifts and cross-nonlinearities. The theory is cross-checked through the observed renormalized Rabi frequencies and energy relaxation times.
We clearly see the breakdowns of the PT and TS model as well as RWA.
%

\textit{Theoretical description—}
The Hamiltonian of a bare transmon reads $\hat{H}_q = 4E_C(\hat{N}-N_g) - E_J\cos\hat{\phi}$ ($E_J \gg E_C$), where $E_C$, $E_J$, and $N_g$ are the charging, Josephson energies, and offset charges.
$\hat{N}$ and $\hat{\phi}$ are the Cooper-pair number and phase operators.
Let us set $E_n$ and $\ket{n}_{q}$ as the n-th eigenvalue and eigenstate of $\hat{H}_{q}$. 
The fundamental transition frequency and self-nonlinearity of the transmon is then given by $\omega_q=E_1-E_0$ and $\chi_{q} =\omega_q - (E_2-E_1)$.
We define $\hat{d}$ as a normalized dipole operator given by $\eta\hat{N}$, where $\eta=-i({32E_C}/{E_J})^{1/4}$ (See \textbf{Supplemental Material A1}).
It is important to note that $\hat{d}$ is an anti-Hermitian operator.

A microwave drive adds an additional term $\hat{H}_d(t) = \Omega_d\hat{d}\sin{\omega_d t}$.
Here, $\Omega_d$ and $\omega_d$ are the drive amplitude and frequency, respectively. 
We then invoke a unitary operator $\hat{U}_q(t)$ that satisfies $\hat{U}_q(t)[\hat{H}_q+\hat{H}_d(t)-i\partial/\partial{t}]\hat{U}^\dagger_q(t) = \hat{K}_q$, where $\hat{K}_q$ denotes an effective static `Kamiltonian', which only captures the slow dynamics of the system.
The $n$-th eigenenergies of $\hat{K}_q$ will be expressed by $\widetilde{E}_n$.
$\hat{K}_q$ should be set such that $\widetilde{E}_n$ is adiabatically connected to $E_n$ as $\Omega_d \rightarrow{0}$.
It is also useful to define the renormalized dipole elements $\widetilde{d}_{nm}^{(\pm)}$, which satisfy (See \textbf{Supplemental Material A1 and A2}).
\begin{equation}
\begin{split}
\label{eq:dressed-dipole}
   \hat{U}_q(t)\hat{d}\hat{U}_q^\dagger(t) \cong  \sum_{n,m} \mp  \widetilde{d}_{nm}^{(\pm)}e^{i(m-n\pm1)\omega_d t} ~\hat{\sigma}_{nm}.
\end{split}
\end{equation}
Here, $\hat{\sigma}_{nm}=\ket{n}_q\bra{m}_q$.
The sign distinguishes whether the elements originally concern the absorption or emission processes.
Eq.~\ref{eq:dressed-dipole} concerns the renormalization of Rabi frequencies and energy relaxation times.

%

The renormalization can also be explored for the interaction between a microwave-dressed transmon and the quantized field of a dispersively coupled readout resonator.
We define $\hat{H}(t)$ like below,
\begin{equation}
\begin{split}
\label{eq:dressed-interaction_1}
   \hat{H}(t) = \hat{H}_q + \hat{H}_d(t) + \underbrace{\omega_r\hat{a}^{\dagger}\hat{a}}_{\hat{H}_r} +\underbrace{g\hat{d}(\hat{a}-\hat{a}^{\dagger})}_{\hat{H}_{I}}.
\end{split}
\end{equation}
Here, $\hat{a}$ and $\omega_r$ denote the field operator and frequency of the resonator. 
$\hat{H}_{I}$ denotes the interaction between the transmon and resonator. $g$ refers to the coupling constant.
The renormalization of the transmon–resonator interaction can be nicely captured by the expression below, 
\begin{equation}
\begin{split}
\label{eq:dressed-interaction_2}
   \hat{U}_q\hat{H}(t)\hat{U}_q^{\dagger} = \hat{K}_q + \omega_r\hat{a}^{\dagger} \hat{a} +\underbrace{g[\hat{U}_q\hat{d}\hat{U}_q^{\dagger}](\hat{a}-\hat{a}^{\dagger})}_{\hat{\widetilde{H}}_I}.
\end{split}
\end{equation}
The effects of the renormalized interaction terms ($\hat{\widetilde{H}}_{I}$) and renormalized bare transmon ($\hat{K}_{q}$) are disentangled in Eq. \ref{eq:dressed-interaction_2} whereas they have been ambiguated in the theoretical descriptions of previous works \cite{multi-7}.
We better define $\hat{\widetilde{H}}_{SI}$ by collecting only the static components of $\hat{\widetilde{H}}_{I}(t)$ to distinguish purely static effects from $\hat{\widetilde{H}}_{I}(t)$.
We also define $\hat{K}$ as an effective static form of $\hat{H}(t)$, related by $\hat{U}(t)$ satisfying $\hat{K} = \hat{U}(t)[\hat{H}(t)-i\partial/\partial{t}]\hat{U}^{\dagger}(t)$.
See \textbf{Supplemental Material A3} for how to derive $\hat{U}(t)$.
\begin{center}
\begin{table}
 \begin{tabular}{||c c||} 
 \hline
 Hamiltonian & Effective static form \\ [0.5ex] 
 \hline\hline
 $\hat{H}_q+\hat{H}_d(t)$ & $\hat{K}_q$\\ 
 \hline
 $\hat{K}_q+\hat{H}_{I}+\hat{H}_{r}$ & $\hat{K}_1$\\ 
 \hline
 $\hat{K}_q+\hat{\widetilde{H}}_{SI}+\hat{H}_{r}$ & $\hat{K}_2$\\ 
 \hline
  $\hat{H}_q+\hat{H}_d(t)+\hat{H}_I+\hat{H}_r$ ( $\hat{K}_q+\hat{\widetilde{H}}_{I}+\hat{H}_{r}$) & $\hat{K}$ \\ 
 \hline
  $\hat{H}_\textsf{TS}+\hat{H}_{d,\textsf{TS}}(t)+\hat{H}_{I,\textsf{TS}}+\hat{H}_r$ & $\hat{K}_\textsf{TS}$\\ 
 \hline
\end{tabular}
\caption{\label{tab1} Hamiltonians and their effective static forms used in this work. See main text for their definitions.}
\end{table}
\end{center}
Tab.~\ref{tab1} summarizes all the Hamiltonian models and their effective static forms used in this work.
$\hat{K}_1$ and $\hat{K}_2$ describe the transmon–resonator system with interaction terms $\hat{H}_{I}$ and $\hat{\widetilde{H}}_{SI}$, respectively. 
%
%
%
We also define a corresponding TS system by
$\hat{H}_\textsf{TS}=\frac{\omega_{0,\textsf{TS}}}{2}\hat{\sigma}_z$. 
Its interaction with the resonator is expressed by $\hat{H}_{I,\textsf{TS}}=g_{\textsf{TS}}\hat{\sigma}_x(\hat{a}+\hat{a}^\dagger)$.

\textit{Experimental results and analysis—}
A graphical description of this system is presented in Fig~\ref{fig1}a, which shows
A dispersively coupled microwave-dressed transmon (blue) and resonator (red).
Fig~\ref{fig1}b shows the energy diagram of $\hat{K}$.
The energy levels are probed through resonator transmission and qubit two-tone spectroscopy. The resonator is used to readout the qubit states in the two-tone spectroscopy. See \textbf{Supplemental Material E} for the details.
Here, $n_{q,r}$ denote the excitation numbers of the transmon and resonator, respectively.
Each horizontal line represents an eigenstate of $\hat{K}$.
$\widetilde{\omega}_{q}^{n_r}$ ($\widetilde{\omega}_{r}^{n_q}$) refers to the transmon (resonator) frequency when the resonator (transmon) is in the $n_r$ ($n_q$) energy state.
We additionally introduce symbols $L_q= \omega_q^{0} - \omega_q$ and $\chi_{qr} =\omega_q^{0} -\omega_q^{1}$, which refer to the Lamb shift and cross-nonlinearity in the fundamental transition of the transmon.
From the two-tone spectroscopy \cite{two-tone,two-tone2}, we observe $\omega_q^{0}/2\pi=5.867$ GHz, $\omega_r^{0}/2\pi=4.289$ GHz, $\chi_q^{0}/2\pi=149$ MHz, 
and $\chi_{qr}/2\pi= 6$ MHz.
From this observation we extract the parameters, $E_J/2\pi =28.6$ GHz, $E_C/2\pi=149$ MHz, $\omega_r/2\pi = 4.334$ GHz, $g/2\pi = 245$ MHz, and $L_q/2\pi=33$ MHz.
Both $\omega_{0,\textsf{TS}}$ and $g_{\textsf{TS}}$ in $\hat{H}_{\textsf{TS}}$ are properly adjusted such that they yield the same $\chi_{qr}$, $\omega_q^{0}$, and $\omega_r^{0}$ compared to those of the transmon.
\begin{figure}
    \centering
    \includegraphics[width=1\columnwidth]{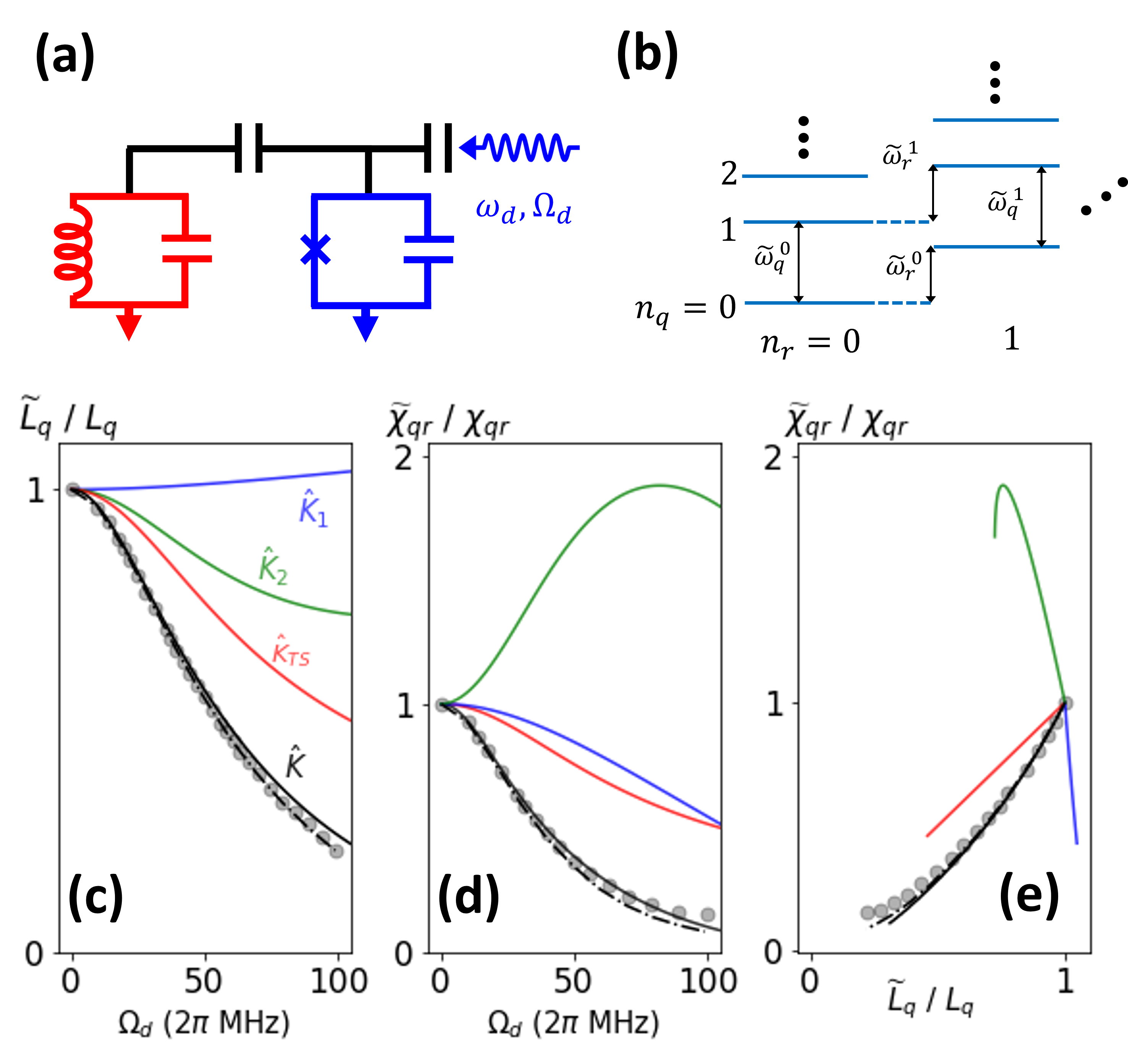}
    \caption{Renormalized interaction between the transmon (blue) and resonator (red) for $\omega_d/2\pi$=5.89 GHz.
    (a,b) Circuit and energy level diagrams. A blue wavy arrow indicates a microwave drive to the transmon.
    (c,d,e) Lamb shift ($\widetilde{L}_{q}$) and cross-nonlinearity ($\widetilde{\chi}_{qr}$) divided by the unnormalized values ($L_q$, $\chi_{qr}$). $\omega_d$ is 5.89 GHz in the experiment. Circles: experimental data. Black, blue, green, red solid lines refer to theoretical calculations based on $\hat{K}$, $\hat{K}_1$, $\hat{K}_2$, and $\hat{K}_{\textsf{TS}}$, respectively. Dashed line refers to fully numerical calculation based on Eq.~\ref{eq:dressed-interaction_1}. See Tab.~\ref{tab1} and the corresponding main text for the description of each model. Statistical errors in data are negligible, and thus not presented in the plots.}
    \label{fig1}
\end{figure}

In Fig.~\ref{fig1}c-e, we present experimentally observed Lamb shifts and cross-nonlinearities of the transmon from two-tone spectroscopy.
$\omega_d/2\pi$ is 5.89 GHz, sufficiently near $\omega_{q}^0/2\pi$.
We explore both renormalized cross-nonlinearity and Lamb shift, denoted by $\widetilde{\chi}_{qr}$ and $\widetilde{L}_{q}$, in Fig.~\ref{fig1}c-e.
$\widetilde{\chi}_{qr}$ alone is not sufficient to fully comprehend the renormalized interaction between the transmon and resonator since it also largely depends on $\widetilde{\chi}_q$.
Since $\widetilde{L}_{q}$ is almost independent of $\widetilde{\chi}_q$, it can be used to investigate the renormalization effects that originate from the transmon–resonator interaction.

The discrepancy between the predictions by the $\hat{K}_1$ model (blue lines) and the experimental data indicates significant renormalization of $\hat{H}_{I}$. This can be translated to the breakdown of the low-order PT, which  assumes negligible renormalization of the eigenbasis.
The difference between the green line ($\hat{K}_{2}$) and experimental data throughout Fig.~\ref{fig1}c–e strongly suggests that dynamical components in $\hat{\widetilde{H}}_I$ have significant effects.
The approximation used in $\hat{K}_{1,2}$ often appears when describing the interaction among quantized modes even in the recent studies \cite{multi-6-1, multi-6, multi-6-3, multi-6-2}. In addition, we separately discuss the effect of the rotating wave approximation in \textbf{Supplemental Material A5.}

Although the transmon–drive detuning $\Delta_{qd}^{0}/2\pi = 23$ MHz is substantially smaller than the self-nonlinearity $\chi_q^{0}/2\pi=149$ MHz, the experimental results deviate remarkably from the predictions based on a driven TS system (red lines).
We confirm that the ratio between $L_q$ and $\chi_{qr}$ is invariant with respect to drive fields as seen in the red lines ($\hat{K}_{\textsf{TS}}$) of Fig.~\ref{fig1}c-e.
This shows negligible dynamical effects for the TS system under drive fields.

For dressed TS systems, dynamical effects are expected to be negligible unless $\widetilde{d}_{10}\sim d_{01}$ or $\omega_d$ meets the matching conditions for two-photon sideband transitions  (\textbf{Supplemental Material A4}).
For transmons, there are possibilities for the fundamental transition to be affected by the dynamical part of $\hat{\widetilde{H}}_{I}$ due to the higher energy levels.
Particularly, the diagonal elements of the renormalized dipole matrix $\widetilde{d}_{nn}$ are the major contribution in the case of near resonant drive fields.
For a TS system, $|\widetilde{d}_{00}|=|\widetilde{d}_{11}|$ always holds, and therefore the dynamical effects originating from these components do not induce any energy level shift. 
For a transmon, however, $|\widetilde{d}_{00}| \neq |\widetilde{d}_{11}|$ due to higher energy levels, and then one should also seriously take the diagonal elements into consideration.
See \textbf{Supplemental Material C} for the calculated $\widetilde{d}_{nn}$.

\begin{figure}
    \centering
    \includegraphics[width=1\columnwidth]{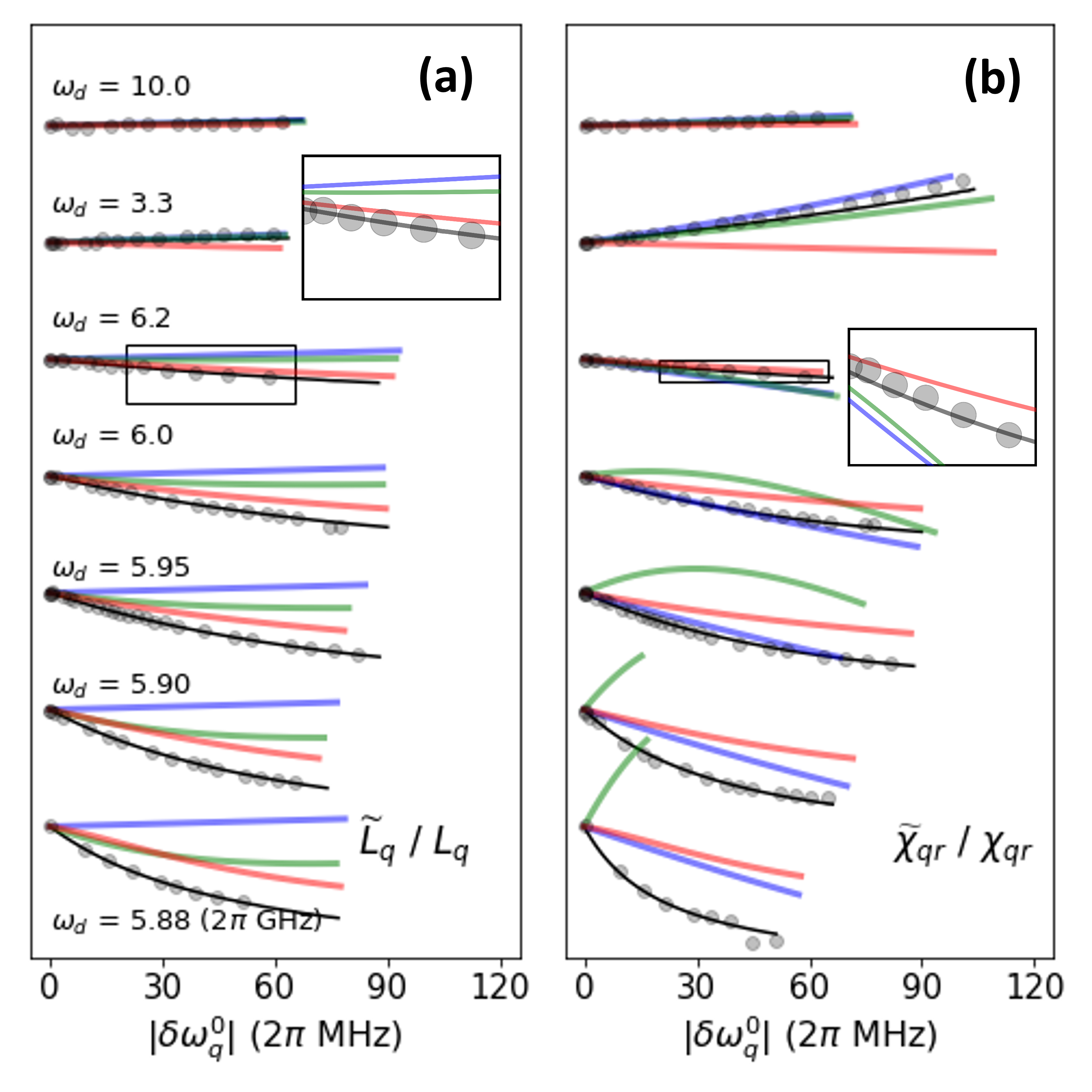}
    \caption{Renormalized interaction between the transmon and resonator with various $\omega_d$. We present renormalized Lamb shifts $\widetilde{L}_q$ and cross-nonlinearities ($\widetilde{\chi}_{qr}$) for given absolute values of Stark shifts ($|\delta\omega_q^0| = |\widetilde{\omega}_q^{0}-\omega_q^{0}|$).
    The circles refer to the observed Lamb shifts and cross-nonlinearities. Theoretical calculations based on several models are denoted by lines. The color legend is identical to that of Fig.~\ref{fig1}. Insets give magnified views of the areas enclosed by the boxes. Various tendencies can be seen with respect to the transmon–drive detunings. See the main text for detailed descriptions. Errors in data are negligible, and thus not presented in the plots.
    }
    \label{fig2}
\end{figure}

In Fig.~\ref{fig2}, we present the observed Lamb shifts (a) and cross-nonlinearities (b) as functions of corresponding Stark shifts ($\delta\omega_q^0 = \widetilde{\omega}_q^{0}-\omega_q^{0}$) for various $\omega_d$ from near to far off-resonant regimes.
The investigated range of drive amplitudes are regulated differently for each $\omega_d$, such that the ranges of $|\delta\omega_q^0/2\pi|$ become similar.
%
%
The theoretical model based on $\hat{K}$ (black lines) agrees well with the experimental values (circles) for all $\omega_d$.
Meanwhile, theories based on $\hat{K}_\textsf{TS}$ (red), $\hat{K}_1$ (blue), and $\hat{K}_2$ (green) fail to explain the experimental data in general.
We do not use any free fitting parameter in all the theoretical plots.

For large detunings ($\omega_d/2\pi$=3.3 and 10 GHz), the $\hat{K}_{1}$ model nicely explains the experiments, suggesting $\hat{H}_{I} \approx \hat{\widetilde{H}}_{I}$.
As $\omega_d$ approaches $\omega_q^{0}$, we can clearly see that the deviation of the $\hat{K}_1$ model from the experimental data grows significantly.
When $\omega_d/2\pi$ is near 6.0 GHz, however, we see an exception for the cross-nonlinearities.
We attribute this coincidence to the dynamical effects.
It is interesting to remark that the data with $\omega_d/2\pi$=3.3 GHz exhibits significant renormalization of the cross-nonlinearities but not of the Lamb shifts.
This suggests that the change in $\widetilde{\chi}_{qr}$ for $\omega_d/2\pi$ = 3.3 GHz mainly originates from the change in $\widetilde{\chi}_{q}$.
It is also beneficial to discuss the discrepancy between the predictions from the two-state model and the experiment results.
For large $|\Delta_{qd}|$ ($\omega_d/2\pi$=3.3 and 10 GHz), both $\widetilde{L}_q$ and $\widetilde{\chi}_{qr}$ are nearly invariant from the TS model, indicating $\widetilde{\hat{H}}_{I,TS} \approx \hat{H}_{I,TS}$ under the investigated range of drive amplitudes.
The discrepancy becomes larger when $|\Delta_{qd}|\rightarrow 0$, and this result is already expected from Fig.~\ref{fig1}.
For $\omega_d/2\pi=5.88$ GHz, the corresponding $\Delta_{qd}/2\pi$ is only 13 MHz, which is tenfold smaller than $\chi_q^{0}$. Nonetheless, we still see the dramatic failure of the $\hat{K}_\textsf{TS}$ model.

\begin{figure}
    \centering
    \includegraphics[width=1\columnwidth]{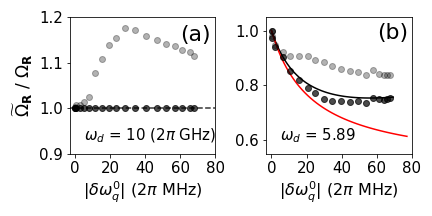}
    \caption{Renormalized Rabi frequencies ($\widetilde{\Omega}_{R}$) for given $|\delta\omega_q^0|$. Here, ${\Omega}_{R}$ indicates the Rabi frequency when $\Omega_d=0$. 
    (a) $\widetilde{\Omega}_{R}/{\Omega}_{R}$ versus $|\delta\omega_q^0|$ for $\omega_d/2\pi=10$ GHz. 
    The gray circles refer to the observed values.  
    When $\Gamma=1$, $\widetilde{\Omega}_{R}\approx{\Omega}_{R}$ should hold for large-detuned drive fields for both the transmon and TS system (dashed line). Based on this, we extract the experimental $\Gamma$, and obtain black circles that correspond to the gray circles.
    (b) $\widetilde{\Omega}_{R}/{\Omega}_{R}$ versus $|\delta\omega_q^0|$ for $\omega_d/2\pi=5.89$ GHz. We divide the observed values (gray circles) by the experimental $\Gamma$ (\new{Eq.~\ref{eq:dressed-rabi}}) obtained from (a), and obtain corrected values with $\Gamma=1$ (black circles).
    The solid lines refer to the approximate theoretical curves under $\Gamma=1$.
    The black and red lines denote transmon and TS system theories, respectively. Errors in data are negligible, and thus not presented in the plots.
    }
    \label{fig3}
\end{figure}


In the following, we investigate the renormalization effects using different approaches, through Rabi frequencies and coherence properties of the transmon. When we introduce an additional Rabi tone that induces resonant transitions $\widetilde{\ket{0}} \leftrightarrow \widetilde{\ket{1}}$ of the dressed transmon, the Rabi frequency $\widetilde{\Omega}_{R}$ satisfies,
\begin{equation}
\begin{split}
\label{eq:dressed-rabi}
   \widetilde{\Omega}_{R} = \underbrace{2\sqrt{\frac{\kappa_{ex}(\widetilde{\omega}_{q}^{0})~\gamma(\widetilde{\omega}_{q}^{0}) ~P_{R}}{\widetilde{\omega}_{q}^{0}}}}_{\Gamma}~\times~\widetilde{d}^{~0(-)}_{01}.
\end{split}
\end{equation}
Here, $\widetilde{d}^{~k(\pm)}_{nm}$ are matrix components of $\hat{U}(t)\hat{d}\hat{U}^\dagger(t)$ when $n_r=k$.
We assume that the frequency of the Rabi tone is adjusted to $\widetilde{\omega}_{q}^{0}$ for resonant Rabi oscillations.
$P_{R}$ is the power of the Rabi tone measured at the output of the microwave source,
$\kappa_{ex}$ refers to the transmon's external coupling to the drive line, and $\Gamma$ is the transfer function between the microwave source and device.
Both $\kappa_{ex}$ and $\Gamma$ are unknown without additional calibrations. 

In Fig.~\ref{fig3}(a), we present $\widetilde{\Omega}_R$ for $\omega_d/2\pi=10$ GHz (gray circles) with respect to absolute values of Stark shifts ($|\delta\omega_q| = |\widetilde{\omega}_q^{0}-\omega_q^{0}|$).
The dashed line refers to the Rabi frequencies when $\Gamma=1$ for both the transmon and TS models.
For such far-off resonant drive fields, under the explored range of AC stark shifts, the renormalization of the transmon's dipole elements is negligible as found in Fig.~\ref{fig2} both for the transmon and TS models.
Hence, the changes in $\widetilde{\Omega}_R$ should be attributed to the changes in $\Gamma$. Based on this, we can extrapolate the experimental $\Gamma$.
By dividing the measured data by the experimental $\Gamma$, we obtain the dark circles in Fig.~\ref{fig3}(a).
In Fig.~\ref{fig3}(b), we explain the data with $\omega_d/2\pi=5.89$ GHz by using the same experimental $\Gamma$ as in Fig.~\ref{fig3}(a).
In Fig.~\ref{fig3}(b), the corrected values (dark circles) show a much better agreement with transmon-based theory than TS theory.
%
%
\begin{figure}
     \centering
     \includegraphics[width=1\columnwidth]{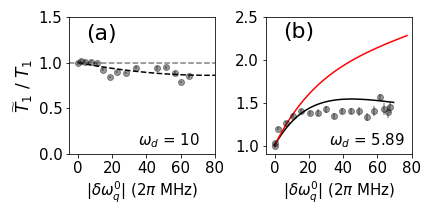}
     \caption{Renormalized energy relaxation time ($\widetilde{T}_1$) for given $|\delta\omega_q^0|$. (a) $\widetilde{T}_1/T_1$ versus $|\delta\omega_q^0|$ is presented for $\omega/2\pi$=10 GHz.  Gray dashed line is a theoretical estimation when  $S_{\lambda_{\perp}}$ is constant with respect to $\delta\omega_q^{0}$.
     The black dashed line is a polynomial fit up to 2nd order. From this, we extrapolate $S_{\lambda_{\perp}}(\widetilde{\omega}_q^0)$. (b) $\widetilde{T}_1/T_1$ versus $|\delta\omega_q^0|$ for $\omega_d/2\pi$=5.89 GHz. Black (transmon) and red (TS) lines refer to theoretical estimations considering non-constant $S_{\lambda_{\perp}}(\widetilde{\omega}_q^0)$.
     }
     \label{fig4}
\end{figure}

The theoretical investigation of renormalization of transmon coherence times is reported recently \cite{multi-14} for sufficiently off-resonant drive fields using perturbative expansion.
In this work, we obtain the non-perturbative solution by applying $\hat{U}_q$ to $\hat{H}_{\textbf{st}}=\lambda_{\parallel}(t)\hat{n}+\lambda_{\perp}(t)\hat{d}$ that describes the interaction between the system and noise environment.
Please recall that we did the same job for $\hat{H}_{I}$ to resolve the renormalized Lamb shift and cross-nonlinearites.
$\hat{n}=\sum_{i}\sqrt{i+1}\ket{i}_q\bra{i}_q$ is the transmon number operator. $\lambda(t)$ is a stochastic variable describing the environmental noise.
We find the relations below (\textbf{See Supplemental Material B}),
\begin{equation}
\begin{split}
\label{eq:dressed-coherence}
   &\frac{1}{\widetilde{T}_1} = \pi \left[S_{{\lambda}_{\perp}}(\widetilde{\omega}_{q}^{0})(\widetilde{d}^{~0(-) 2}_{01}) \right], \\
   &\frac{1}{\widetilde{T}_{\varphi }} = \pi\left [S_{{\lambda}_{\parallel }}(0)  (\widetilde{n}^{0}_{11}-\widetilde{n}^{0}_{00})^2+S_{{\lambda}_{\perp}}(\omega_d)(\widetilde{d}_{11}^{~0(-)}-\widetilde{d}^{~0(-)}_{00})^2\right].
\end{split}
\end{equation}
$\widetilde{T}_{2}$ is the renormalized pure dephasing time satisfying $\widetilde{T}_2^{-1} = (2\widetilde{T}_1)^{-1} + \widetilde{T}_{\varphi}^{-1}$.
Here, $S_{\lambda}(\omega)=\frac{1}{2\pi}\int d\tau e^{-i \omega\tau}\left \langle \lambda^*(t)\lambda(t+\tau) \right \rangle $.
$\widetilde{n}^{k}_{nm}$ are the matrix elements of $\hat{\widetilde{n}} = \hat{U}\hat{n}\hat{U}^\dagger$, when $n_r=k$.
Only the fluctuations in the transmon's resonant frequency are taken into account in this expression.
Eq.~\ref{eq:dressed-coherence} becomes identical to Eq. 42 and 45 in \cite{Dressed-tls-15} in the two-state approximation.

In Fig.~\ref{fig4}(a), we present $\widetilde{T}_1/T_1$ with respect to $|\delta\omega_q^0|$ for $\omega_d/2\pi=10$ GHz (circles).
The theoretical estimation under the ideal situation ($S_{\lambda_{\perp}}=\textbf{const.}$) is given by the gray dashed line. 
The flatness of the gray line is due to the fact that $\widetilde{d}_{01}^{~0(-)} \approx {d}_{01}^{~0(-)}$ should hold for far-off resonant drive fields.
Thereby, we can draw a conclusion that the change in $\widetilde{T}_1/T_1$ is attributed to the $\omega_q^{0}$ dependence of $S_{\lambda_{\perp}}$.
Therefore, we can extrapolate $S_{\lambda_{\perp}}(\widetilde{\omega}_q^{0})/S_{\lambda_{\perp}}(\omega_q^{0})$ by low-order polynomial fitting (dark dashed-line).
Based on this extrapolation, we fit the data with $\omega_d/2\pi=5.89$ GHz in Fig.~\ref{fig4}(b).
Unfortunately, we cannot thoroughly verify the second equation of Eq.~\ref{eq:dressed-coherence} due to experimental limitations.
See \textbf{Supplemental Material F} for our investigation on the renormalized dephasing times ($\widetilde{T}_{2}$).
See \textbf{Supplemental Material G} for the calculated $\widetilde{\Omega}_{R}$, $\widetilde{T}_{1}$, and $\widetilde{T}_{2}$ based on our formula with various drive frequencies.

\textit{conclusion—}
To summarize, we have verified the non-perturbative renormalization of a coupled transmon–resonator system.
The significant renormalization of the transmon–resonator interaction is identified from the changes of Lamb shifts and cross-nonlinearities.
The results are also consistent with the renormalized Rabi frequencies and energy relaxation times observed separately.
Without using recursive formulas and Floquet theory, we quantitatively explain the experiments.
Our work represents a significant step from the previous relevant works confined to single-mode or two-state descriptions.

Although the performed experiments are confined to a weakly anharmonic circuit QED system, overall strategies to account for the renormalization will concern a broad range of cavity QED-like systems or more generally even to coupled multi-mode systems.
Furthermore, transmons are also well-known examples of Duffing oscillators or pendulums in quantum regimes, and thus our work will also contribute to fundamental understanding on driven nonlinear systems.

%
%


\begin{acknowledgements}
We thank David Thoen and Jochem Baselmans for providing us with NbTiN films. Byoung-moo Ann acknowledges support from the European Union’s Horizon 2020 research and innovation program under the Marie Sklodowska-Curie grant agreement No. 722923 (OMT).  Byoung-moo Ann also acknowledges support from Korea Research Institute of Standards and Science (KRISS-2023-GP2023-0012-22) and Basic Science Research Program through the National Research Foundation of Korea (NRF) funded by the Ministry of Education (NP2023-0007).
Please see Supplemental Material \cite{sm} for detailed information on the theoretical derivation, numerical simulation method, device, experimental setup, and data acquisition process.
Data supporting the plots within this paper are available through Zenodo at \cite{data}.
Further information is available from the corresponding author upon reasonable request.
\end{acknowledgements}


\widetext
\clearpage


\renewcommand{\theequation}{S.\arabic{equation}}
\renewcommand{\thepage}{S\arabic{page}} 
\renewcommand{\thesection}{S\arabic{section}}  
\renewcommand{\thetable}{S\arabic{table}}  
\renewcommand{\thefigure}{S\arabic{figure}}
\setcounter{figure}{0}
\setcounter{equation}{0}
\setcounter{table}{0}


\maketitle

\section{A. Derivation of Kamiltonians}
\label{analytical}

\subsection{A1. Microwave-dressed transmon}

In this section, we will derive the effective static Hamiltonians for a bare transmon subject to a monochromatic microwave drive.
The drive Hamiltonian is given by
$\hat{H}_d(t) = i\eta\Omega_d\hat{N}\sin{\omega_d t}$.
The microwave-dressed transmon has an analogy with a driven charged particle on a cosine potential.
$\hat{N}$ and $\hat{\phi}$ correspond to momentum and position, respectively.
Considering the velocity and length gauge equivalence, $\hat{H}_d(t)$ can be also expressed as $\hat{H}_d(t) = \zeta\Omega_d\hat{\phi}\cos{\omega_d t}$, where $\zeta=\omega_d/\sqrt{8E_JE_c}$.
Here, $\zeta\Omega_d$ can be considered as the drive amplitude defined on the other gauge.
For convenience in the derivation, we use $\hat{H}_d(t) = \zeta\Omega_d\hat{\phi}\cos{\omega_d t}$ from now on.

\begin{figure}
    \centering
    \includegraphics[width=1.0\columnwidth]{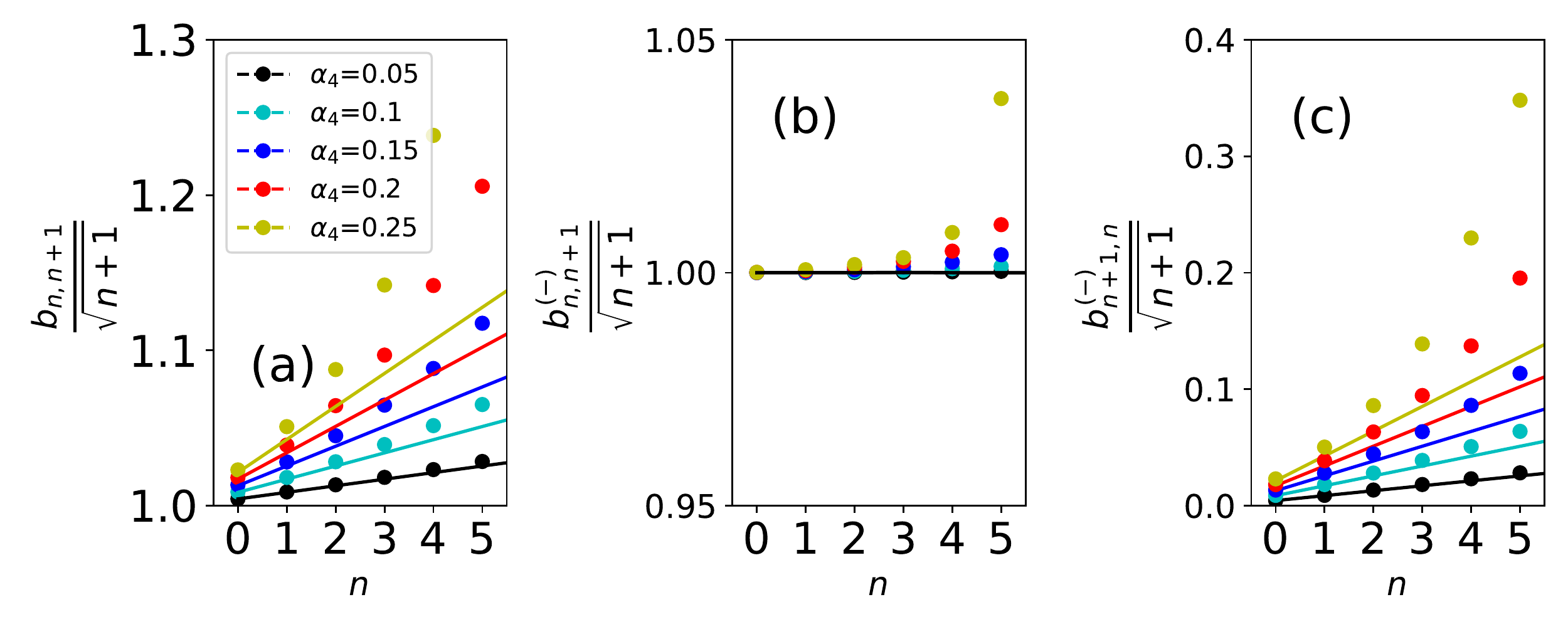}
    \caption{Elements of $\hat{U}_q^{(2)}\hat{b}\hat{U}_q^{(2)\dagger}$ and $\hat{U}_q^{(2)}\hat{b}^\dagger\hat{U}_q^{(2)\dagger}$ on the eigenbasis of $\hat{H}_q$ for various $\alpha_4$. 
    The lines are analytical results with only the leading order effect. The circles are results obtained by numerically calculating  $\hat{U}_q^{(2)}$.
    }
    \label{sm-b}
\end{figure}

Then, we introduce a ladder operator $\hat{b} = \frac{1}{\sqrt{2}}(\frac{E_J}{8E_C})^{\frac{1}{4}}\hat{\phi} + i\frac{1}{\sqrt{2}}(\frac{8E_C}{E_J})^{\frac{1}{4}}\hat{N}$, and recast the Hamiltonian by
\begin{equation}
\begin{split}
\label{eq:1}
  \hat{H}_q &= (\overline{\omega} + \alpha_4)\hat{b}^\dagger\hat{b} - \frac{\alpha_4}{12}(\hat{b}+\hat{b}^\dagger)^4 + \sum_{n=3}^{\infty}\alpha_{2n}(\hat{b}+\hat{b}^\dagger)^{2n}. \\
  \hat{H}_d &= \zeta{\Omega}_d(\hat{b}+\hat{b}^\dagger)\cos{\omega_d t}.
\end{split}
\end{equation}
Here, we set $N_g=0$. For transmons, the effect of $N_g$ is negligible unless the drive is strong enough to induce the ionization of the ground and first excited states \cite{multi-11-sm, multi-12-sm}. In this work, we confine ourselves to the drive regime where the effects of $N_g$ are negligible.
The difference between $\omega_q$ and $\overline{\omega} + \alpha_4$ is caused by the off-diagonal elements in the nonlinear terms of $\hat{H}_q$.
The first hurdle we need to overcome to get the effective static Hamiltonian $\hat{K}_q$ are the counter-rotating (CR) terms in $\hat{H}_d$.
We take a unitary transformation $\hat{U}_q^{(1)}=e^{i\xi\hat{X}}$, $\hat{X}=\frac{\zeta\Omega_d}{\overline{\omega}_q} \sin \omega_d t (\hat{b}+\hat{b}^\dagger)$ to eliminate those terms. 
Here, $\xi$ is a general complex number that will be determined later.
For simplicity, we define $\overline{\omega}_q=\overline{\omega}+\alpha_4$, and $\hat{H}_q^{(0)} = \hat{H}_q+\hat{H}_d$.
Taking the transformation $\hat{U}_q^{(1)}$ yields
\begin{equation}
\begin{split}
\label{eq:2}
  \hat{H}_q^{(1)} & = \hat{U}_q^{(1)}[\hat{H}_q^{(0)}-i\partial_t]\hat{U}_q^{(1)\dagger} \\ & = \overline{\omega}_q\hat{b}^\dagger\hat{b} - \frac{\alpha_4}{12}(\hat{b}+\hat{b}^\dagger)^4 + \sum_{n=3}^{\infty}\alpha_{2n}(\hat{b}+\hat{b}^\dagger)^{2n}+(1-\xi)\zeta\Omega_d\cos\omega_d t(\hat{b}+\hat{b}^\dagger)+i\zeta\xi\Omega_d\frac{\overline{\omega}_q}{\omega_d}\sin\omega_d t(\hat{b}-\hat{b}^\dagger).
\end{split}
\end{equation}
Here, we use the Baker–Campbell–Hausdorff (BCH) formula, $e^{\xi\hat{X}}\hat{H}e^{-\xi\hat{X}} = \hat{H} - \xi\left [ \hat{H}, \hat{X} \right ] + \frac{\xi^2}{2} \left[  \left[\hat{H} , \hat{X} \right] , \hat{X} \right] \cdots $.
Since $\left[ \hat{H}_{q}^{(0)}-\overline{\omega}_q\hat{b}^\dagger\hat{b}  , \hat{X} \right]=0$, the calculations can be significantly simplified. 
Setting $\xi=(1+\frac{\overline{\omega}_q}{\omega_d})^{-1}$, then we can eliminate all the counter-rotating terms in the drive components of Eq.~\ref{eq:2},
\begin{equation}
\begin{split}
\label{eq:3}
  \hat{H}_q^{(1)} =  \overline{\omega}_q\hat{b}^\dagger\hat{b} - \frac{\alpha_4}{12}(\hat{b}+\hat{b}^\dagger)^4 + \sum_{n=3}^{\infty}\alpha_{2n}(\hat{b}+\hat{b}^\dagger)^{2n}+\frac{\overline{\Omega}_d}{2}(e^{i\omega_d t}\hat{b}+e^{-i\omega_d t}\hat{b}^\dagger),
\end{split}
\end{equation}
where, $\overline{\Omega}_d = \zeta\Omega_d(1-\xi)+\zeta\xi\Omega_d(\overline{\omega}_q/\omega_d)$.
Note that we have not yet introduced any approximations when going from Eq~\ref{eq:1} to Eq.~\ref{eq:3}.
The next step is to remove the off-diagonal terms in the static part of Eq.~\ref{eq:3}. Let us define a unitary operator $\hat{U}_q^{(2)}$ that satisfies
\begin{equation}
\label{eq:4}
  \hat{H}_q^{(2)} = \hat{U}_q^{(2)}\hat{H}_q^{(1)}\hat{U}_q^{(2)\dagger} =  \omega_q\hat{b}^\dagger\hat{b} - \frac{a_4}{2}\hat{b}^{\dagger 2}\hat{b}^2 + \sum_{n=3}^{\infty}a_{2n}\hat{b}^{\dagger n}\hat{b}^{n}+\frac{\overline{\Omega}_d}{2}\hat{U}_q^{(2)}(e^{i\omega_d t}\hat{b}+e^{-i\omega_d t}\hat{b}^\dagger)\hat{U}_q^{(2)\dagger}.
\end{equation}
The static part in Eq.~\ref{eq:4} is now fully diagonalized, but we still need works on the drive part.
We decompose $\hat{U}_q^{(2)}\hat{b}\hat{U}_q^{(2)\dagger}$ and $\hat{U}_q^{(2)}\hat{b}^{\dagger}\hat{U}_q^{(2)\dagger}$ below like,
\begin{equation}
\begin{split}
\label{eq:5}
    \hat{U}_q^{(2)}\hat{b}\hat{U}_q^{(2)\dagger} & = \sum_{n>m} b_{nm}^{(+)}\ket{n}\bra{m} + \sum_{n<m} b_{nm}^{(-)}\ket{n}\bra{m} \\
    \hat{U}_q^{(2)}\hat{b}^{\dagger}\hat{U}_q^{(2)\dagger} & = \sum_{n>m} b^{(+)}_{mn}\ket{n}\bra{m} + \sum_{n<m} b^{(-)}_{mn}\ket{n}\bra{m}.
\end{split}
\end{equation}
Here, $\ket{n}$ indicates the eigenstates of $\hat{H}_q$ or $\hat{H}^{(1)}_q$ when $\overline{\Omega}_d=0$. 
We also define $b_{nm} = b^{(+)}_{nm} + b^{(+)}_{mn} = b^{(-)}_{nm} + b^{(-)}_{mn}$.
$b_{nn}$ are always zero in general and not presented in Eq.~\ref{eq:5}.
The signs indicate whether the corresponding elements regard absorption or emission processes.
Before taking the next steps, let us investigate which terms in Eq.~\ref{eq:5} are negligible.
First, we only keep the elements satisfying $|n-m|=1$, neglecting  harmonic generation processes.
This occurs when $\omega_d \sim (2k+1)\omega_q$ and $k \in \textbf{N}$, which is rarely satisfied in the typical cases.
We numerically confirm the validity of this assumption in Fig.~\ref{sm-e}.
Thereby, the drive term in Eq.~\ref{eq:4} can be approximated as
\begin{equation}
\begin{split}
\label{eq:6}
     & \hat{U}_q^{(2)}(e^{i\omega_d t}\hat{b}+e^{-i\omega_d t}\hat{b}^\dagger)\hat{U}_q^{(2)\dagger} \approx \\
    & \sum_{n} \left (\cancelto{\approx0}{b_{n+1,n}^{(+)}\ket{n+1}\bra{n}e^{i\omega_d t}} + b_{n,n+1}^{(-)}\ket{n}\bra{n+1}e^{i\omega_d t} + b^{(+)}_{n,n+1}\ket{n+1}\bra{n}e^{-i\omega_d t} + \cancelto{\approx0}{b^{(-)}_{n+1,n}\ket{n}\bra{n+1}e^{-i\omega_d t}}\right).
\end{split}
\end{equation}
In Fig.~\ref{sm-b}, we present the calculated $b_{n,n+1}$, $b_{n,n+1}^{(-)}$, and $b_{n+1,n}^{(-)}$, normalized by $\sqrt{n+1}$. In the calculation, we set $\alpha_{2n} = 0$ for $n>2$ since the contribution from the higher-order terms is negligible to the results.
We can confirm that $b_{n,n+1}^{(-)} \gg b_{n+1,n}^{(-)}$ in Fig.~\ref{sm-b}, and it is subsequently expected that $b_{n,n+1}^{(+)} \gg b_{n+1,n}^{(+)}$ is satisfied.
Fortunately, the coefficients of CR terms ($b_{n+1,n}^{(-)}$,$b_{n+1,n}^{(+)}$) of the right-hand side of Eq.~\ref{eq:6} are much smaller than those of the co-rotating terms ($b_{n,n+1}^{(-)}$,$b_{n,n+1}^{(+)}$)
unless $n \gg 1$.
As long as $\frac{b_{n,n+1}^{(-)}}{\omega_q - \omega_d},\frac{b_{n,n+1}^{(+)}}{\omega_q - \omega_d}$ are much larger than $\frac{b_{n+1,n}^{(-)}}{\omega_q + \omega_d}, \frac{b_{n+1,n}^{(+)}}{\omega_q + \omega_d}$, we can neglect the CR terms (cancelled-out) in Eq.~\ref{eq:6}.
These conditions nicely hold for small $n$.
When describing the interaction between the qubit and resonator, the conventional rotating wave approximation (RWA) requires $\frac{\Sigma_{qd}}{\Delta_{qd}} \gg 1$, 
where $\omega_q - \omega_d = \Delta_{qd}$ and $\omega_q + \omega_d = \Sigma_{qd}$.
The approximation made in Eq.~\ref{eq:6} requires 
$\frac{\Sigma_{qd}}{\Delta_{qd}} \frac{b_{n,n+1}^{(\pm)}}{b_{n+1,n}^{(\pm)}} \gg 1$
Since we have already proven $\frac{b_{n,n+1}^{(\pm)}}{b_{n+1,n}^{(\pm)}} \gg 1$ in Fig.~\ref{sm-b} for small $n$,
Eq.~\ref{eq:6} is a better approximation than the conventional rotating wave approximation (RWA).

Therefore, $\hat{H}_q^{(2)}$ can be approximated by
\begin{equation}
\begin{split}
\label{eq:7}
  \hat{H}_q^{(2)} \approx  \omega_q\hat{b}^\dagger\hat{b} - \frac{a_4}{2}\hat{b}^{\dagger 2}\hat{b}^2 + \sum_{n=3}^{\infty}a_{2n}\hat{b}^{\dagger n}\hat{b}^{2n}+\frac{\overline{\Omega}_d}{2}\sum_{n}(b^{(-)}_{n,n+1}\ket{n}\bra{n+1}e^{i\omega_d t} + b^{(+)}_{n,n+1}\ket{n+1}\bra{n}e^{-i\omega_d t}).
\end{split}
\end{equation}
By taking $\hat{U}_q^{(3)}=e^{-i\sum_{n} n\omega_d t\ket{n}\bra{n}}$ on $\hat{H}_q^{(2)}$, we eventually obtain the time-independent form, 
\begin{equation}
\begin{split}
\label{eq:8}
  \hat{H}_q^{(3)} \approx \hat{U}_q^{(3)}[\hat{H}_q^{(2)}-i\partial_t]\hat{U}_q^{(3)\dagger} = (\omega_q-\omega_d)\hat{b}^\dagger\hat{b} - \frac{a_4}{2}\hat{b}^{\dagger 2}\hat{b}^2 + \sum_{n=3}^{\infty}a_{2n}\hat{b}^{\dagger n}\hat{b}^{n}+\frac{\overline{\Omega}_d}{2}\sum_{n}(b^{(-)}_{n,n+1}\ket{n}\bra{n+1} + b^{(+)}_{n,n+1}\ket{n+1}\bra{n}).
\end{split}
\end{equation}
$\hat{H}_q^{(3)}$ can be readily diagonalized without any additional methods to cope with the challenges of time-dependent problems.
We define $\hat{U}_q^{(4)}$ that diagonalizes $\hat{H}_q^{(3)}$, satisfying the equation below:
\begin{equation}
\begin{split}
\label{eq:9}
  \hat{H}_q^{(4)} \approx \hat{U}_q^{(4)}\hat{H}_q^{(3)}\hat{U}_q^{(4)\dagger} = (\widetilde{\omega}_q-\omega_d)\hat{b}^{\dagger}\hat{b} - \frac{\widetilde{a}_4}{2}\hat{b}^{\dagger 2}\hat{b}^2 + \sum_{n=3}^{\infty}\widetilde{a}_{2n}\hat{b}^{\dagger n}\hat{b}^{n}.
\end{split}
\end{equation}
This step can be performed analytically without recursive approach in principle. 
In this work, however, we rely on QuTip \cite{Qutip1, Qutip2} to find $\hat{H}_q^{(4)}$ to save time.
$\hat{H}_q^{(4)}$ is not yet the final form since its eigenspectra is not adiabatically connected to $\hat{H}_q$ as $\Omega_d \rightarrow 0$. 
In order to satisfy this condition, we should take one more transformation $\hat{U}_q^{(5)}= \hat{U}_q^{(3) -1} = e^{i\sum_{n} n\omega_d t \ket{n}\bra{n}}$.
Finally, we come to a conclusion of this section,
\begin{equation}
\begin{split}
\label{eq:10}
  \hat{K}_q \approx \hat{U}_q^{(5)}[\hat{H}_q^{(4)}-i\partial_t]\hat{U}_q^{(5)\dagger} = \widetilde{\omega}_q\hat{b}^\dagger\hat{b} - \frac{\widetilde{a}_4}{2}\hat{b}^{\dagger 2}\hat{b}^2 + \sum_{n=3}^{\infty}\widetilde{a}_{2n}\hat{b}^{\dagger n}\hat{b}^{n}.
\end{split}
\end{equation}
Please note that $\hat{U}_q^{(4)}$ is the only transformation that practically mixes the eigenstates, and the others just concern the phase of the states.
Finally, we can approximately express $\hat{U}_q$ by
\begin{equation}
\begin{split}
\label{eq:11}
    \hat{U}_q \approx \hat{U}_q^{(5)}\hat{U}_q^{(4)}\hat{U}_q^{(3)}\hat{U}_q^{(2)}\hat{U}_q^{(1)}.
\end{split}
\end{equation}

\subsection{A2. Renormalized dipole of microwave-dressed transmons}

In this section, we will verify Eq.~1 of the main text.
With ladder operators, the normalized dipole operator $\hat{d}$ is $\hat{b} - \hat{b}^\dagger$.
The left-hand side the equation can be expressed by
\begin{equation}
\begin{split}
\label{eq:2-1}
    \hat{U}_q\hat{d}\hat{U}_q^\dagger=\hat{U}_q(\hat{b}-\hat{b}^\dagger)\hat{U}_q^\dagger \approx (\hat{U}_q^{(5)}\hat{U}_q^{(4)}\hat{U}_q^{(3)}\hat{U}_q^{(2)}\hat{U}_q^{(1)})(\hat{b}-\hat{b}^\dagger)(\hat{U}_q^{(1)\dagger}\hat{U}_q^{(2)\dagger}\hat{U}_q^{(3)\dagger}\hat{U}_q^{(4)\dagger}\hat{U}_q^{(5)\dagger})
\end{split}
\end{equation}
First, taking $\hat{U}_q^{(1)}$ on $\hat{b}-\hat{b}^\dagger$ just yields the constant shifts by $i\xi\hat{I}$, where $\hat{I}$ is the identity operator. 
Thereby, we can neglect $\hat{U}_q^{(1)}$ in Eq.~\ref{eq:2-1}.   
The result after taking the second unitary transformation $\hat{U}_q^{(2)}$ can be readily obtained from Eq.~\ref{eq:5},
\begin{equation}
\begin{split}
\label{eq:2-2}
    \hat{d}^{(2)} = \hat{U}_q^{(2)}\hat{d}\hat{U}_q^{(2)\dagger} =\hat{U}_q^{(2)}(\hat{b}-\hat{b}^\dagger)\hat{U}_q^{(2)\dagger} = \sum_{n<m} \underbrace{(b_{nm}^{(-)}+b_{mn}^{(-)})}_{d^{(-)}_{nm}}\ket{n}\bra{m} - \sum_{n>m} \underbrace{(b_{nm}^{(+)}+b_{mn}^{(+)})}_{d^{(+)}_{nm}}\ket{n}\bra{m} , 
\end{split}
\end{equation}
where, $d^{(\pm)}_{nm} = b_{nm}^{(\pm)}+b_{mn}^{(\pm)}$.
Subsequently taking $\hat{U}_q^{(3)}$ yields
\begin{equation}
\begin{split}
\label{eq:2-3}
    \hat{d}^{(3)} = \hat{U}_q^{(3)}\hat{d}^{(2)}\hat{U}_q^{(3)\dagger} = \sum_{n<m} {d^{(-)}_{nm}}\ket{n}\bra{m}e^{-i(n-m)\omega_d t} - \sum_{n>m} d^{(+)}_{nm}\ket{n}\bra{m}e^{i(n-m)\omega_d t}. 
\end{split}
\end{equation}
As decided in Eq.~\ref{eq:6}, we neglect the terms not satisfying $|n-m|=1$, and this gives
\begin{equation}
\begin{split}
\label{eq:2-4}
    \hat{d}^{(3)} \approx  \sum_{n} {d^{(-)}_{n,n+1}}\ket{n}\bra{n+1}e^{-i\omega_d t} - \sum_{n} d^{(+)}_{n+1,n}\ket{n+1}\bra{n}e^{i\omega_d t}. 
\end{split}
\end{equation}
After taking $\hat{U}_q^{(4)}$, we obtain the following:
\begin{equation}
\begin{split}
\label{eq:2-5}
    \hat{d}^{(4)} = \hat{U}_q^{(4)}\hat{d}^{(3)}\hat{U}_q^{(4)\dagger} \approx \sum_{n} {d^{(-)}_{n,n+1}}\widetilde{\ket{n}}\widetilde{\bra{n+1}}e^{-i\omega_d t} - \sum_{n} {d^{(+)}_{n+1,n}}\widetilde{\ket{n+1}}\widetilde{\bra{n}}e^{i\omega_d t},
\end{split}
\end{equation}
where $\widetilde{\ket{n}} = \hat{U}_q^{(4)}\ket{n}$.
Using $\hat{I} = \sum_{n}\ket{n}\bra{n}$, we can recast Eq.~\ref{eq:2-5}  in the form
\begin{equation}
\begin{split}
\label{eq:2-6}
    \hat{d}^{(4)} \approx  \sum_{n,i,j} {d^{(-)}_{n,n+1}}\langle{i}\widetilde{\ket{n}}\widetilde{\bra{n+1}}{j}\rangle\ket{i}\bra{j}e^{-i\omega_d t} - \sum_{n,i,j} d^{(+)}_{n+1,n}\langle  {i}\widetilde{\ket{n+1}}\widetilde{\bra{n}}{j}\rangle\ket{i}\bra{j}e^{i\omega_d t}.
\end{split}
\end{equation}
Now, we define the following:
\begin{equation}
\begin{split}
\label{eq:2-7}
    \widetilde{d}^{(+)}_{ij} = \sum_{n} d^{(+)}_{n+1,n}\langle  {i}\widetilde{\ket{n+1}}\widetilde{\bra{n}}{j}\rangle,\\
    \widetilde{d}^{(-)}_{ij} = \sum_{n} d^{(-)}_{n,n+1}\langle  {i}\widetilde{\ket{n}}\widetilde{\bra{n+1}}{j}\rangle.
\end{split}
\end{equation}
Then, Eq.~\ref{eq:2-6} can be expressed as,
\begin{equation}
\begin{split}
\label{eq:2-8}
    \hat{d}^{(4)} \approx \sum_{n,m} \widetilde{d}_{nm}^{(-)}\ket{n}\bra{m}e^{-i\omega_d t}-\sum_{n,m} \widetilde{d}_{nm}^{(+) }\ket{n}\bra{m}e^{i\omega_d t}.
\end{split}
\end{equation}
Beware that we replace $i,j$ with $n,m$ for consistency with the main text.
Taking $\hat{U}_q^{(5)}$ finally yields the same expression as Eq.~1 in the main text:
\begin{equation}
\begin{split}
\label{eq:2-9}
    \hat{U}_q\hat{d}\hat{U}_q^\dagger\ \approx \hat{U}_q^{(5)}\hat{d}^{(4)}\hat{U}_q^{(5)\dagger} \approx \sum_{n,m} \widetilde{d}_{nm}^{(-)}\ket{n}\bra{m}e^{i(m-n-1)\omega_d t} - \sum_{n,m} \widetilde{d}_{nm}^{(+) }\ket{n}\bra{m}e^{i(m-n+1)\omega_d t}.
\end{split}
\end{equation}

\subsection{A3. Microwave-dressed transmons dispersively coupled to a quantized field}
With ladder operators $\hat{b}$ and $\hat{b}^\dagger$, the Hamiltonian of the microwave-dressed transmon dispersively coupled to a resonator ($\hat{H}$) is expressed by
\begin{equation}
\begin{split}
\label{eq:3-1}
  \hat{H} &= \overline{\omega}_q\hat{b}^\dagger\hat{b} - \frac{\alpha_4}{12}(\hat{b}+\hat{b}^\dagger)^4 + \sum_{n=3}^{\infty}\alpha_{2n}(\hat{b}+\hat{b}^\dagger)^{2n} +\omega_r\hat{a}^\dagger\hat{a}+g(\hat{a} - \hat{a}^\dagger)(\hat{b}-\hat{b}^\dagger)  + \zeta{\Omega}_d(\hat{b}+\hat{b}^\dagger)\cos{\omega_d t}.
\end{split}
\end{equation}
Here, $g$ is the coupling constant between the transmon and resonator and $\omega_r$ is the frequency of the resonator.
We take unitary transformations $\hat{U}^{(1)}=\hat{U}_q^{(1)}$, and then obtain the equation below
\begin{equation}
\begin{split}
\label{eq:3-2}
  \hat{H}^{(1)} = \hat{U}^{(1)}(\hat{H}-i\partial_t)\hat{U}^{(1)\dagger} &= \overline{\omega}_q\hat{b}^\dagger\hat{b} - \frac{\alpha_4}{12}(\hat{b}+\hat{b}^\dagger)^4 + \sum_{n=3}^{\infty}\alpha_{2n}(\hat{b}+\hat{b}^\dagger)^{2n} +\omega_r\hat{a}^\dagger\hat{a}+g(\hat{a} - \hat{a}^\dagger)(\hat{b}-\hat{b}^\dagger) \\ &+ i\frac{g\zeta\Omega_d}{\overline{\omega}_q}\sin \omega_d t(\hat{a} - \hat{a}^\dagger) + \frac{\overline{\Omega}_d}{2}(e^{i\omega_d t}\hat{b}+e^{-i\omega_d t}\hat{b}^\dagger).
\end{split}
\end{equation}
Here, $\overline{\Omega}_d = \zeta\Omega_d(1-\xi)+\zeta\xi\Omega_d(\overline{\omega}_q/\omega_d)$, and $\xi$ is the same as in Eq.~\ref{eq:2}.
We then apply a displacement operator, $\hat{U}^{(2)}=e^{\xi_2^{*}(t)\hat{a}-\xi_2(t)\hat{a}^\dagger}$, where
$\xi_2(t)=\frac{g\zeta\Omega_{d}}{2\overline{\omega}_q\Delta_{rd}}e^{-i\omega_{d}t}-\frac{g\zeta\Omega_{d}}{2\overline{\omega}_q\Sigma_{rd}}e^{i\omega_{d}t}$.
Here, $\Delta_{rd}$ and $\Sigma_{rd}$ refer to $\omega_r - \omega_d$ and $\omega_r + \omega_d$, respectively.
This yields
\begin{equation}
\begin{split}
\label{eq:3-3}
  \hat{H}^{(2)} = \hat{U}^{(2)}(\hat{H}^{(1)}-i\partial_t)\hat{U}^{(2)\dagger} &= \overline{\omega}_q\hat{b}^\dagger\hat{b} - \frac{\alpha_4}{12}(\hat{b}+\hat{b}^\dagger)^4 + \sum_{n=3}^{\infty}\alpha_{2n}(\hat{b}+\hat{b}^\dagger)^{2n}+\omega_r\hat{a}^\dagger\hat{a} +g(\hat{a} - \hat{a}^\dagger) (\hat{b}-\hat{b}^\dagger)\\ &+\underbrace{\frac{\overline{\Omega}_d}{2}(e^{i\omega_d t}\hat{b}+e^{-i\omega_d t}\hat{b}^\dagger) + i\left ( \frac{g^2\zeta\Omega_d}{\overline{\omega}_q\Delta_{rd}}-\frac{g^2\zeta\Omega_d}{\overline{\omega}_q\Sigma_{rd}} \right )\sin\omega_d t(\hat{b}-\hat{b}^\dagger)}_{\approx\frac{1}{2}\left (\overline{\Omega}_d+\frac{g^2\zeta\Omega_d}{\overline{\omega}_q\Delta_{rd}}\right)e^{i\omega_d t}\hat{b} + \frac{1}{2}\left (\overline{\Omega}_d+\frac{g^2\zeta\Omega_d}{\overline{\omega}_q\Delta_{rd}}\right)e^{-i\omega_d t}\hat{b}^\dagger -\cancelto{\approx0}{\frac{g^2\zeta\Omega_d}{2\overline{\omega}_q\Delta_{rd}}e^{i\omega_d t}\hat{b}^\dagger}-\cancelto{\approx0}{\frac{g^2\zeta\Omega_d}{2\overline{\omega}_q\Delta_{rd}}e^{-i\omega_d t}\hat{b}}}.
\end{split}
\end{equation}
We have removed the CR terms (cancelled-out) in the original drive Hamiltonian, but obtained additional CR components. 
Fortunately, we can neglect these CR terms that newly show up as long as $\overline{\Omega}_d \gg \frac{g^2\zeta\Omega_d}{2\overline{\omega}_q|\Delta_{rd}|}$ and $\frac{\Sigma_{qd}}{\Delta_{qd}} \gg \frac{g^2\zeta}{2\overline{\omega}_q|\Delta_{rd}|}$.
We also do not need to consider any accidental resonant interactions induced by CR terms.
This approximation is much tighter than the conventional RWA.
With redefining $\overline{\Omega}_d \rightarrow \overline{\Omega}_d+\frac{g^2\zeta\Omega_d}{\overline{\omega}_q\Delta_{rd}}$, Eq.~\ref{eq:3-3} can be approximated as
\begin{equation}
\begin{split}
\label{eq:3-4}
  \hat{H}^{(2)} \approx \overline{\omega}_q\hat{b}^\dagger\hat{b} - \frac{\alpha_4}{12}(\hat{b}+\hat{b}^\dagger)^4 + \sum_{n=3}^{\infty}\alpha_{2n}(\hat{b}+\hat{b}^\dagger)^{2n} +\omega_r\hat{a}^\dagger\hat{a}+g(\hat{a}-\hat{a}^\dagger)(\hat{b} - \hat{b}^\dagger) +\frac{\overline{\Omega}_d}{2}e^{i\omega_d t}\hat{b}+\frac{\overline{\Omega}_d}{2}e^{-i\omega_d t}\hat{b}^\dagger.
\end{split}
\end{equation}
We now apply $\hat{U}^{(3)}=e^{\xi_3\hat{X}_3}$, where $\hat{X}_3 = \hat{a}^\dagger\hat{b}^\dagger - \hat{a}\hat{b}$ and $\xi_3 = g/(\overline{\omega}_q+\omega_r)$ to eliminate the CR terms in the transmon–resonator interaction Hamiltonian.
Then, the result is presented below:
\begin{equation}
\begin{split}
\label{eq:3-5}
  & \hat{H}^{(3)} = \hat{U}^{(3)}\hat{H}^{(2)}\hat{U}^{(3) \dagger}  \approx \\ &\underbrace{(\overline{\omega}_q-2g\xi_3)\hat{b}^\dagger\hat{b} - \frac{\alpha_4}{12}(\hat{b}+\hat{b}^\dagger-\xi_3\hat{a}-\xi_3\hat{a}^\dagger)^4 + \sum_{n=3}^{\infty}\alpha_{2n}(\hat{b}+\hat{b}^\dagger-\xi_3\hat{a}-\xi_3\hat{a}^\dagger)^{2n} + ({\omega}_r-2g\xi_3)\hat{a}^\dagger\hat{a} + g\xi_3(\hat{a}^2+\hat{a}^{\dagger2}+\hat{b}^2+\hat{b}^{\dagger2}) + \hat{O}( \xi_3^2)}_{\hat{H}^{(3)}_0} \\ & \underbrace{- g(\hat{a}\hat{b}^\dagger + \hat{a}^\dagger\hat{b}) + \frac{\overline{\Omega}_d}{2}e^{i\omega_d t}\hat{b}+\frac{\overline{\Omega}_d}{2}e^{-i\omega_d t}\hat{b}^\dagger-\cancelto{\approx0}{\frac{\overline{\Omega}_d\xi_3}{2}e^{i\omega_d t}\hat{a}}- \cancelto{\approx0}{\frac{\overline{\Omega}_d^{*}\xi_3}{2}e^{-i\omega_d t}\hat{a}^\dagger} + \cancelto{\approx0}{\hat{O}_{d}(\xi_3^2)}}_{\hat{H}^{(3)}_1}.
\end{split}
\end{equation}
For convenience in the following discussion, we decompose $\hat{H}^{(3)}$ into $\hat{H}^{(3)}_0$ and $\hat{H}^{(3)}_1$.
Here, we define $\hat{O}_d(\xi_3^2)$ as the collection of second or higher order terms of $\xi_3$ in the drive Hamiltonian. For these terms, we have $\hat{O}_d(\xi_3^2) \rightarrow 0$ as $\Omega_d \rightarrow 0$.
In $\hat{H}^{(3)}_1$, the terms proportional to $\hat{a}$ and $\hat{a}^\dagger$ can be absorbed in the other terms proportional to $\hat{b}$ and $\hat{b}^\dagger$ by applying a displacement operator, as in Eq.~\ref{eq:3-3}.
This leads to some time-dependent terms and a slight renormalization of $\overline{\Omega}_d$, but we can neglect these effects as long as $1 \gg \frac{g\xi_3}{|\Delta_{rd}|}$ hold.
We also set $\hat{O}_d(\xi_3^2)\approx 0$ unless $\omega_d$ meets specific matching conditions for resonant interactions.
However, we will keep $\hat{O}(\xi_3^2)$ in $\hat{H}_0^{(3)}$, since the diagonal components of $\hat{O}$ can still make noticeable contributions.

As the next step, we apply $\hat{U}^{(4)}$ on $\hat{H}_0^{(3)}$, which satisfies
\begin{equation}
\begin{split}
\label{eq:3-6}
  \hat{H}^{(4)}_0 = \hat{U}^{(4)}\hat{H}^{(3)}_0\hat{U}^{(4) \dagger} = \overline{\omega}_{q}^0\hat{b}^\dagger\hat{b} + \overline{\omega}_{r}^0\hat{a}^\dagger\hat{a}  - \frac{a_{4,0}}{2}\hat{a}^{\dagger 2}\hat{a}^2 - a_{2,2}\hat{a}^{\dagger}\hat{a}\hat{b}^{\dagger}\hat{b} - \frac{a_{0,4}}{2}\hat{b}^{\dagger 2}\hat{b}^2+ \sum_{n+m=3}^{\infty}a_{2n,2m}\hat{a}^{\dagger n}\hat{a}^{n}\hat{b}^{\dagger m}\hat{b}^{m}.
\end{split}
\end{equation}
In practice, we extract $\overline{\omega}_{q,r}^0$ and $a_{n,m}$ from the experimental results obtained by two-tone spectroscopy.
In this work, we neglect $a_{2n, 2m}$ for $n+m>3$, and confirm that higher order terms with $n+m>3$ negligibly contribute to the lower energy levels.
Then, we make the following approximation, 
\begin{equation}
\begin{split}
\label{eq:3-6-1}
  & \hat{H}^{(4)}_1 = \hat{U}^{(4)}\hat{H}^{(3)}_1\hat{U}^{(4) \dagger} \approx \hat{U}_q^{(4)}\hat{H}^{(3)}_1\hat{U}_q^{(4) \dagger}.
\end{split}
\end{equation}
This step can be justified when the added diagonal terms in $\hat{H}^{(3)}_0$ which converge to zero as $\xi_3\rightarrow0$ are perturbative, such that these terms induce noticeable corrections to eigenenergies, but negligible corrections to eigenstates.
Then, the drive terms become
\begin{equation}
\begin{split}
\label{eq:3-7}
   \hat{H}^{(4)}_1 \approx -g\sum_{n}(b^{(-)}_{n,n+1}\hat{a}\ket{n}\bra{n+1} +b^{(+)}_{n,n+1}\hat{a}^\dagger\ket{n+1}\bra{n}) + \frac{\overline{\Omega}_d}{2}\sum_{n}b^{(-)}_{n,n+1}\ket{n}\bra{n+1}e^{i\omega_d t} + \frac{\overline{\Omega}_d}{2}\sum_{n}b^{(+)}_{n,n+1}\ket{n+1}\bra{n}e^{-i\omega_d t}.
\end{split}
\end{equation}
By taking $\hat{U}^{(5)}=e^{-i\sum_{n} n\omega_d t\ket{n}\bra{n} }e^{-i\omega_d t\hat{a}^\dagger\hat{a}}$ on $\hat{H}_0^{(4)}+\hat{H}_1^{(4)}$, we eventually obtain the time-independent form $\hat{H}^{(5)}$ like below,
\begin{equation}
\begin{split}
\label{eq:3-7}
    &\hat{H}^{(5)} \approx \hat{U}^{(5)}(\hat{H}^{(4)}-i\partial_t)\hat{U}^{(5)\dagger} = (\overline{\omega}_{q}^0-\omega_d)\hat{b}^\dagger\hat{b} + (\overline{\omega}_{r}^0-\omega_d)\hat{a}^\dagger\hat{a}  - \frac{a_{4,0}}{2}\hat{a}^{\dagger 2}\hat{a}^2 - a_{2,2}\hat{a}^{\dagger}\hat{a}\hat{b}^{\dagger}\hat{b} - \frac{a_{0,4}}{2}\hat{b}^{\dagger 2}\hat{b}^2 \\& + \sum_{n+m=3}^{\infty}a_{2n,2m}\hat{a}^{\dagger n}\hat{a}^{n}\hat{b}^{\dagger m}\hat{b}^{m} 
    -g\sum_{n}(b^{(-)}_{n,n+1}\hat{a}^\dagger\ket{n}\bra{n+1} +b^{(+)}_{n,n+1}\hat{a}\ket{n+1}\bra{n}) \\& + \frac{\overline{\Omega}_d}{2}\sum_{n}b^{(-)}_{n,n+1}\ket{n}\bra{n+1} + \frac{\overline{\Omega}_d}{2}\sum_{n}b^{(+)}_{n,n+1}\ket{n+1}\bra{n}.
\end{split}
\end{equation}
Here, $\hat{H}^{(4)} = \hat{H}_0^{(4)}+\hat{H}_1^{(4)}$.
Let us define $\hat{U}^{(6)}$ that diagonalizes $\hat{H}^{(5)}$. We also define $\hat{H}^{(6)}=\hat{U}^{(6)}\hat{H}^{(5)}\hat{U}^{(6)\dagger}$. As we did in Sec.~A1, we apply $\hat{U}^{(7)} = [\hat{U}^{(5)}]{}^{-1}$ to make the eigenenergies are adiabatically connected to the undressed Hamiltonian as $\Omega_d \rightarrow 0$.
Finally, we obtain $\hat{K}$ like below, 
\begin{equation}
\begin{split}
\label{eq:3-7}
    &\hat{K} \approx \hat{U}^{(7)}(\hat{H}^{(6)}-i\partial_t)\hat{U}^{(7)\dagger} =  \widetilde{\omega}_{q}^{0}\hat{b}^\dagger\hat{b} + \widetilde{\omega}_{r}^{0}\hat{a}^\dagger\hat{a}  - \frac{\widetilde{\chi}_r^0}{2}\hat{a}^{\dagger 2}\hat{a}^2 - \widetilde{\chi}_{qr}\hat{a}^{\dagger}\hat{a}\hat{b}^{\dagger}\hat{b} - \frac{\widetilde{\chi}_{q}^0}{2}\hat{b}^{\dagger 2}\hat{b}^2+ \sum_{n+m=3}^{\infty}\widetilde{a}_{2n,2m}\hat{a}^{\dagger n}\hat{a}^{n}\hat{b}^{\dagger m}\hat{b}^{m}.
\end{split}
\end{equation}
Here, a newly defined symbol $\tilde{\chi}_{r}^n$ refers to the inherited self-nonlinearity of the resonator when the transmon is in $\ket{n}$ state. 

\subsection{A4. Two-state systems}

A time-periodically driven two-state system dispersively coupled to a resonator mode can be expressed by,
\begin{equation}
\begin{split}
\label{eq:4-1}
    \hat{H}_\textsf{TS}+\hat{H}_{d,\textsf{TS}}+\hat{H}_{I,\textsf{TS}}+\hat{H}_r=\frac{\omega_{0,\textsf{TS}}}{2}\hat{\sigma}_z+\Omega_d\hat{\sigma}_x\cos\omega_d t + g_{\textsf{TS}}\hat{\sigma}_x(\hat{a}+\hat{a}^\dagger)+\omega_r \hat{a}^\dagger\hat{a}.
\end{split}
\end{equation}
Individual definitions of the terms on the LHS of Eq.~\ref{eq:4-1} are given in the main text.
For TS systems, we can use the counter-rotating hybridized rotating wave approximation (CHRW) discussed in \cite{CHRH1,CHRH2} to approximate $\hat{H}_\textsf{TS}+\hat{H}_{d,\textsf{TS}}(t)$ like below:
\begin{equation}
\begin{split}
\label{eq:4-1}
    \hat{H}_\textsf{TS}+\hat{H}_{d,\textsf{TS}} \approx \frac{\overline{\omega}_{0,\textsf{TS}}}{2}\hat{\sigma}_z + \frac{\overline{\Omega}_d}{2}\left(\hat{\sigma}^{-}e^{i\omega_d t} + \hat{\sigma}^{+}e^{-i\omega_d t} \right).
\end{split}
\end{equation}
Here, $\overline{\omega}_{0,\textsf{TS}} = \omega_{0,\textsf{TS}}J_0(2\Omega_d\xi_{\textsf{TS}}/\omega_d)$ and $\overline{\Omega}_d=2\omega_{0,\textsf{TS}}J_1\left(\frac{2\Omega_d}{\omega_d}\xi_{\textsf{TS}} \right)$. $\xi_{\textsf{TS}}$ is given such that it satisfies $\Omega_d(1-\xi_{\textsf{TS}})=\omega_{0,\textsf{TS}}J_1\left(\frac{2\Omega_d}{\omega_d}\xi_{\textsf{TS}} \right)$. $J_n$ refer to Bessel functions of the first kind.
The effective static Hamiltonian of Eq.~\ref{eq:4-1} can be calculated straightforwardly.
We introduce $\hat{U}_{\textsf{TS}}(t)$ that satisfies
\begin{equation}
\begin{split}
\label{eq:4-2}
    \hat{U}_{\textsf{TS}}(t)\ket{e} = e^{i\omega_d t}\sin{\frac{\theta}{2}}\ket{g} + \cos{\frac{\theta}{2}}\ket{e}\\
    \hat{U}_{\textsf{TS}}(t)\ket{g} = \cos{\frac{\theta}{2}}\ket{g} - e^{-i\omega_d t}\sin{\frac{\theta}{2}}\ket{e}.
\end{split}
\end{equation}
The dressed TS system's Hamiltonian is then transformed to,
\begin{equation}
\begin{split}
\label{eq:4-3}
    \hat{U}_{\textsf{TS}}(t)(\hat{H}_\textsf{TS}+\hat{H}_{d,\textsf{TS}}-i\partial_t)\hat{U}_{\textsf{TS}}^\dagger(t) = \frac{\overline{\omega}_{0,\textsf{TS}}+\delta}{2}\hat{\sigma}_z.
\end{split}
\end{equation}
Here, $\delta = \sqrt{(\overline{\omega}_{0,\textsf{TS}}-\omega_d)^2+\Omega_d^2}$ and $\theta=\texttt{arctan}(\Omega_d/(\overline{\omega}_{0,\textsf{TS}}-\omega_d))$.
The interaction term $\hat{H}_{I,\textsf{TS}}$ is transformed to
\begin{equation}
\begin{split}
\label{eq:4-4}
    \hat{U}_{\textsf{TS}}(t)\hat{H}_{I,\textsf{TS}}\hat{U}_{\textsf{TS}}^\dagger(t) = g_{\textsf{TS}}(\cos^2\frac{\theta}{2}\hat{\sigma}_x - e^{2i\omega_d t}\sin^2\frac{\theta}{2}\hat{\sigma}^{-} -e^{-2i\omega_d t}\sin^2\frac{\theta}{2}\hat{\sigma}^{+} - 2\cos{\omega_d t}\cos{\frac{\theta}{2}}\sin{\frac{\theta}{2}}\hat{\sigma}_z)(\hat{a}+\hat{a}^\dagger).
\end{split}
\end{equation}
When $\omega_d$ is off-resonant to both $\omega_r$ and $\omega_{0,\textsf{TS}}$, the longitudinal diagonal terms with rotating speed $\omega_d$ in Eq.~\ref{eq:4-4} do not yield any frequency shifts of the systems when considering only the leading order effect.
The other possibility that can account for frequency shifts is two-photon sideband transition terms with rotating speed $2\omega_d$.
Unless we have $\theta \sim \pi/2$, or $\omega_d$ sharply meets the matching conditions for the two-photon sideband transitions, their effects are negligible compared with those of static interaction term in Eq.~\ref{eq:4-4}.
Therefore, except for some specific conditions listed here, Eq.~\ref{eq:4-5} can be approximated as
\begin{equation}
\begin{split}
\label{eq:4-5}
    \hat{U}_{\textsf{TS}}(t)\hat{H}_{I,\textsf{TS}}\hat{U}_{\textsf{TS}}^\dagger(t) \approx g_{\textsf{TS}}(\cos^2\frac{\theta}{2}\hat{\sigma}_x)(\hat{a}+\hat{a}^\dagger).
\end{split}
\end{equation}
Renormalizing the interaction term approximately amounts to replacing the coupling constant $g_{\textsf{TS}}$ with $g_{\textsf{TS}}\cos^2\frac{\theta}{2}$.
This explains why the ratio between the Lamb shift and cross-nonlinearity is almost conserved with increasing drive amplitudes.

\subsection{A5. Effects of counter-rotating terms.}

The rotating wave approximation (RWA) neglects the counter-rotating terms in the Hamiltonian, and consequently, makes the problem of finding the static Kamiltonian tremendously simplified. Although it has been widely adopted in various systems, it often breaks down in circuit QED. In this section, we present the results under the RWA and compare them to our derivation in the previous sections. 
Capturing only co-rotating terms in $\hat{H}_q$, $\hat{H}_d$, and $\hat{H}$ yields $\hat{H}_q^\textsf{{rwa}}$, $\hat{H}_d^\textsf{{rwa}}$, and $\hat{H}^\textsf{{rwa}}$.
\begin{equation}
\begin{split}
\label{eq:5-1}
  \hat{H}_q^\textsf{{rwa}} &= \overline{\omega}\hat{b}^\dagger\hat{b} - \frac{\alpha_4^\textsf{{rwa}}}{2}\hat{b}^\dagger\hat{b}^\dagger\hat{b}\hat{b} + \sum_{n=3}^{\infty}\alpha_{2n}^\textsf{{rwa}}\hat{b}^{\dagger n}\hat{b}^{n}. \\
  \hat{H}_d^\textsf{{rwa}} &= \frac{\zeta{\Omega}_d}{2}(\hat{b}e^{i\omega_d t}+\hat{b}^\dagger e^{-i\omega_d t}).
\end{split}
\end{equation}
\begin{equation}
\begin{split}
\label{eq:5-1-1}
  \hat{H}^\textsf{{rwa}} = \overline{\omega}_q\hat{b}^\dagger\hat{b} - \frac{\alpha_4^\textsf{{rwa}}}{2}\hat{b}^\dagger\hat{b}^\dagger\hat{b}\hat{b} + \sum_{n=3}^{\infty}\alpha_{2n}^\textsf{{rwa}}\hat{b}^{\dagger n}\hat{b}^{n} +\omega_r\hat{a}^\dagger\hat{a}-g(\hat{a}\hat{b}^\dagger + \hat{a}^\dagger\hat{b}) + \frac{\zeta{\Omega}_d}{2}(\hat{b}e^{i\omega_d t}+\hat{b}^\dagger e^{-i\omega_d t}).
\end{split}
\end{equation}
Let us take an unitary transform $\hat{U}_q^\textsf{{rwa(1)}} = e^{-i\omega_d t \hat{b}^{\dagger}\hat{b}}$ on $\hat{H}_q^\textsf{{rwa}}+\hat{H}_d^\textsf{{rwa}}(t)$, which results in
\begin{equation}
\begin{split}
\label{eq:5-2}
    \hat{H}_q^\textsf{{rwa(2)}} = \hat{U}_q^\textsf{{rwa(1)}}(t)[\hat{H}_q^\textsf{{rwa}}+\hat{H}_d^\textsf{{rwa}}(t)-i\partial/\partial{t}]\hat{U}^{\textsf{{rwa}}(1)\dagger}_q(t) = (\overline{\omega}_q-\omega_d)\hat{b}^\dagger\hat{b} - \frac{\alpha_4^\textsf{{rwa}}}{2}\hat{b}^\dagger\hat{b}^\dagger\hat{b}\hat{b} + \sum_{n=3}^{\infty}\alpha_{2n}^\textsf{{rwa}}\hat{b}^{\dagger n}\hat{b}^{n} + \frac{\zeta{\Omega}_d}{2}(\hat{b} +\hat{b}^\dagger).
\end{split}
\end{equation}
$\hat{H}_q^\textsf{{rwa(2)}}$ is time-independent, and can be diagonalized by a time-independent unitary transformation $\hat{U}_q^\textsf{{rwa(2)}}$ as below
\begin{equation}
\begin{split}
\label{eq:5-2}
    \hat{H}_q^\textsf{{rwa(2)}} = \hat{U}_q^\textsf{{rwa(2)}}\hat{H}_q^\textsf{{rwa(1)}}\hat{U}^{\textsf{{rwa}}(2)\dagger}_q = (\widetilde{\omega}_q^\textsf{{rwa}}-\omega_d)\hat{b}^\dagger\hat{b} - \frac{\widetilde{a}_4^\textsf{{rwa}}}{2}\hat{b}^\dagger\hat{b}^\dagger\hat{b}\hat{b} + \sum_{n=3}^{\infty}\widetilde{a}_{2n}^\textsf{{rwa}}\hat{b}^{\dagger n}\hat{b}^{n}.
\end{split}
\end{equation}
Taking one more transformation $\hat{U}_q^\textsf{{rwa(3)}}(t)=[\hat{U}_q^\textsf{{rwa(1)}}(t)]^{-1}$ on $\hat{H}_q^\textsf{{rwa(2)}}$ finally yields the static Kamiltonian $\hat{K}_q^\textsf{{rwa}}$ as below
\begin{equation}
\begin{split}
\label{eq:5-2}
    \hat{K}_q^\textsf{{rwa}} = \hat{U}_q^\textsf{{rwa(3)}}(t)[\hat{H}_q^\textsf{{rwa(2)}}-i\partial/\partial{t}]\hat{U}^{\textsf{{rwa}}(3)\dagger}_q(t) = \widetilde{\omega}_q^\textsf{{rwa}}\hat{b}^\dagger\hat{b} - \frac{\widetilde{a}_4^\textsf{{rwa}}}{2}\hat{b}^\dagger\hat{b}^\dagger\hat{b}\hat{b} + \sum_{n=3}^{\infty}\widetilde{a}_{2n}^\textsf{{rwa}}\hat{b}^{\dagger n}\hat{b}^{n}.
\end{split}
\end{equation}

We can find the static Kamiltonian for $\hat{H}^\textsf{{rwa}}$ with a similar approach. We take a unitary operator $\hat{U}^\textsf{{rwa}(1)}(t)=e^{-i\omega_dt\hat{b}^{\dagger}\hat{b}} e^{-i\omega_dt\hat{a}^{\dagger}\hat{a}}$
on $\hat{H}^\textsf{{rwa}}$
\begin{equation}
\begin{split}
\label{eq:5-3}
    \hat{H}^\textsf{{rwa}(1)} & = \hat{U}^\textsf{{rwa(1)}}(t)[\hat{H}^\textsf{{rwa}}-i\partial/\partial{t}]\hat{U}^{\textsf{{rwa}}(1)\dagger}(t) = (\overline{\omega}_q-\omega_d)\hat{b}^\dagger\hat{b} - \frac{\alpha_4^\textsf{{rwa}}}{2}\hat{b}^\dagger\hat{b}^\dagger\hat{b}\hat{b}\\ + &\sum_{n=3}^{\infty}\alpha_{2n}^\textsf{{rwa}}\hat{b}^{\dagger n}\hat{b}^{n} +\omega_r\hat{a}^\dagger\hat{a}-g(\hat{a}\hat{b}^\dagger + \hat{a}^\dagger\hat{b}) + \frac{\zeta{\Omega}_d}{2}(\hat{b}+\hat{b}^\dagger).
\end{split}
\end{equation}
$\hat{H}^\textsf{{rwa(1)}}$ can be diagonalized by a time-independent unitary transformation $\hat{U}^\textsf{{rwa(2)}}$ as below
\begin{equation}
\begin{split}
\label{eq:5-4}
    \hat{H}^\textsf{{rwa}(2)}  = \hat{U}^\textsf{{rwa(2)}}\hat{H}^\textsf{{rwa}(1)}\hat{U}^{\textsf{{rwa}}(2)\dagger} = (\widetilde{\omega}_{q}^{0,\textsf{rwa}}-\omega_d)\hat{b}^\dagger\hat{b} & +  \widetilde{\omega}_{r}^{0,\textsf{rwa}}\hat{a}^\dagger\hat{a} -  \frac{\widetilde{\chi}_r^{0,\textsf{rwa}}}{2}\hat{a}^{\dagger 2}\hat{a}^2 - \widetilde{\chi}_{qr}^\textsf{{rwa}}\hat{a}^{\dagger}\hat{a}\hat{b}^{\dagger}\hat{b} \\ & - \frac{\widetilde{\chi}_{q}^{0,\textsf{rwa}}}{2}\hat{b}^{\dagger 2}\hat{b}^2 + \sum_{n+m=3}^{\infty}\widetilde{a}^\textsf{{rwa}}_{2n,2m}\hat{a}^{\dagger n}\hat{a}^{n}\hat{b}^{\dagger m}\hat{b}^{m}.
\end{split}
\end{equation}
Then, the static Kamiltonian $\hat{K}^\textsf{{rwa}}$ for $\hat{H}^\textsf{{rwa}}$ is given by
\begin{equation}
\begin{split}
\label{eq:5-5}
    \hat{K}^\textsf{{rwa}}  = \hat{U}^\textsf{{rwa(3)}}[\hat{H}^\textsf{{rwa}(2)}-i\partial/\partial{t}]\hat{U}^{\textsf{{rwa}}(3)\dagger} = \widetilde{\omega}_{q}^{0,\textsf{rwa}}\hat{b}^\dagger\hat{b} & +  \widetilde{\omega}_{r}^{0,\textsf{rwa}}\hat{a}^\dagger\hat{a} -  \frac{\widetilde{\chi}_r^{0,\textsf{rwa}}}{2}\hat{a}^{\dagger 2}\hat{a}^2 - \widetilde{\chi}_{qr}^\textsf{{rwa}}\hat{a}^{\dagger}\hat{a}\hat{b}^{\dagger}\hat{b}  \\ & - \frac{\widetilde{\chi}_{q}^{0,\textsf{rwa}}}{2}\hat{b}^{\dagger 2}\hat{b}^2 + \sum_{n+m=3}^{\infty}\widetilde{a}^\textsf{{rwa}}_{2n,2m}\hat{a}^{\dagger n}\hat{a}^{n}\hat{b}^{\dagger m}\hat{b}^{m}.
\end{split}
\end{equation}
Here, $\hat{U}^\textsf{{rwa(3)}} = [\hat{U}^\textsf{{rwa(1)}}]^{-1}$.

\begin{figure}
    \centering
    \includegraphics[width=0.6\columnwidth]{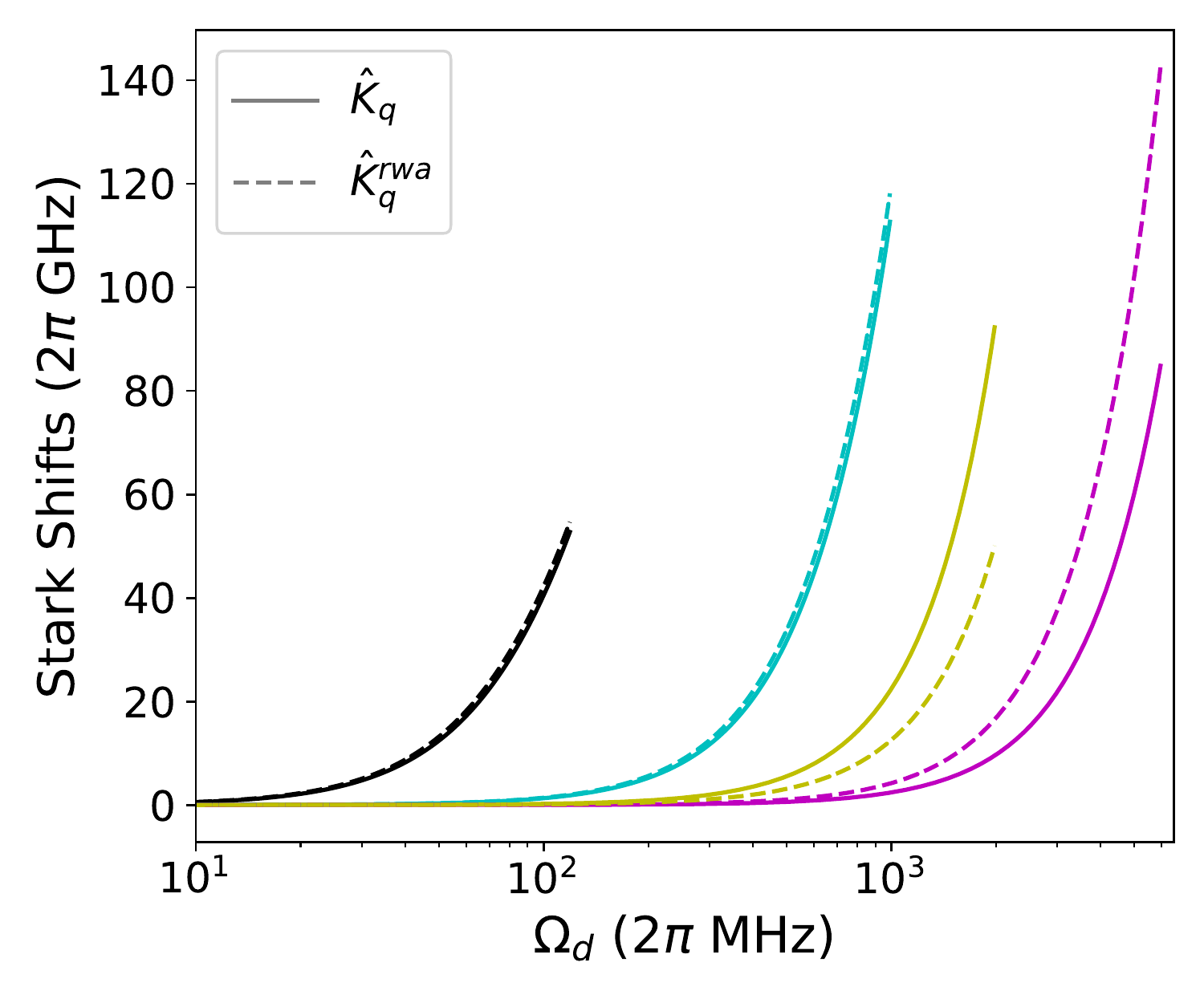}
    \caption{Discrepancy between $\hat{K}_q$  and $\hat{K}_{q}^\textsf{{rwa}}$ model for estimating the Stark shifts from the ground to first eigenstate for various drive frequencies. $\omega_d/2\pi$ for black, cyan, yellow, and magenta are 5.89, 6.5, 3.3, and 10.0 GHz, respectively.
    }
    \label{sm-c}
\end{figure}

\begin{figure}
    \centering
    \includegraphics[width=0.8\columnwidth]{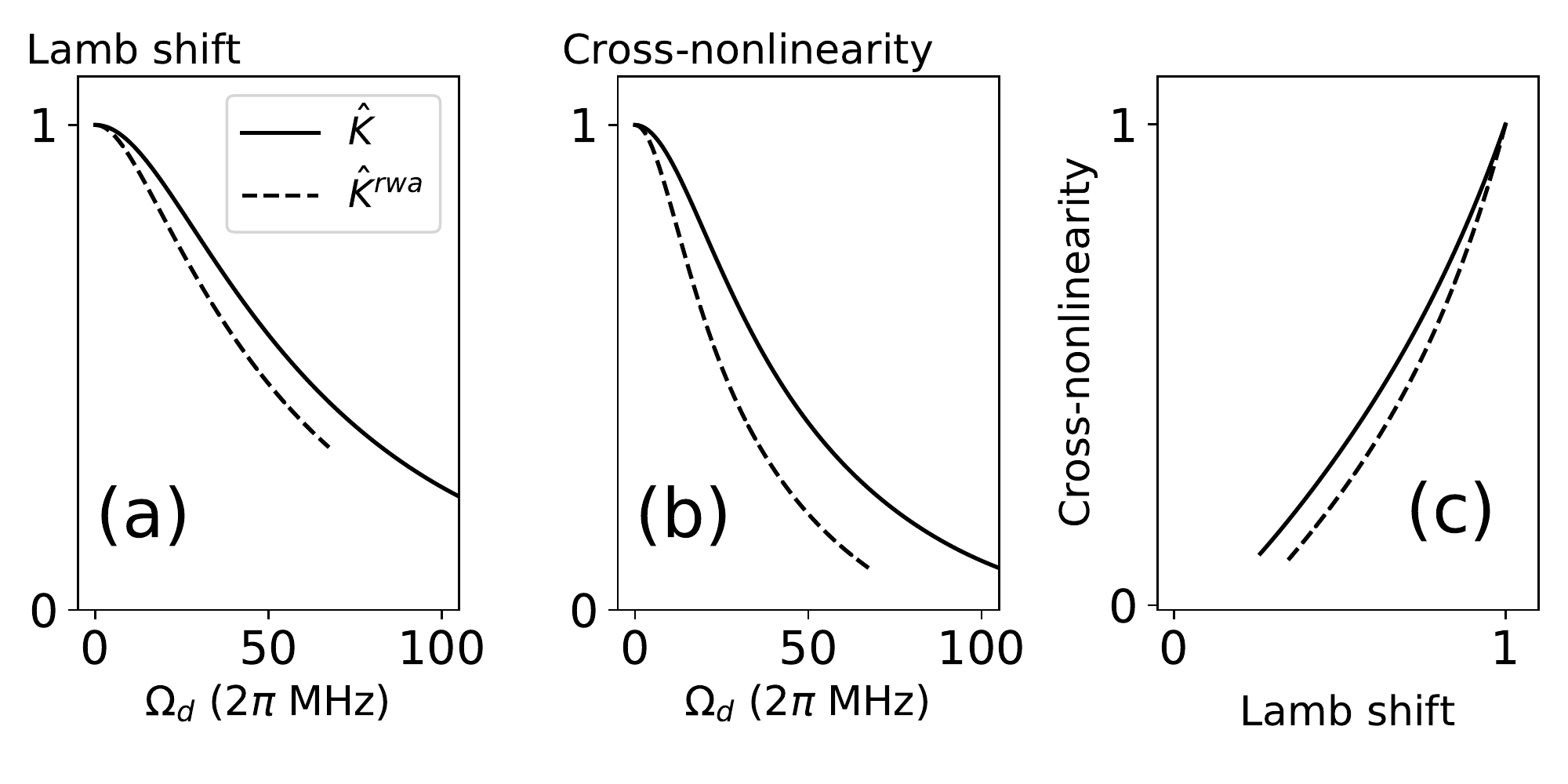}
    \caption{Discrepancy between $\hat{K}$ and $\hat{K}^\textsf{{rwa}}$ model for estimating Lamb shift ($\tilde{L}_q/{L}_q$ or $\tilde{L}_q^\textsf{{rwa}}/{L}_q^\textsf{{rwa}}$) and cross-nonlinearity ($\tilde{\chi}_{qr}/{\chi}_{qr}$ or $\tilde{\chi}_{qr}^\textsf{{rwa}}/{\chi}_{qr}^\textsf{{rwa}}$). $\omega_d/2\pi$ in the calculation is set by 5.89 GHz as in main text figure 1.
    }
    \label{sm-d}
\end{figure}

The renormalized Lamb shift calculated based on the RWA is then given by $\widetilde{L}_q^{\textsf{rwa}}=\widetilde{\omega}_{q}^{0,\textsf{rwa}} - \widetilde{\omega}_{q}^{\textsf{rwa}}$. The renormalized cross-nonlinearity calculated by the RWA is $\widetilde{\chi}_{q}^{0,\textsf{rwa}}$.
In Fig.~\ref{sm-c}, we compare the Stark shifts of the transition frequency from the ground to first eigenstates calculated based on each model. The discrepancy between the $\hat{K}_q$ and $\hat{K}_{q}^\textsf{{rwa}}$ models for far off-resonant transmon–drive detuning (yellow and magenta lines) indicates the breakdown of the RWA.
In Fig.~\ref{sm-d}, we present the Lamb shifts and cross-nonlinearities calculated by the $\hat{K}$ and $\hat{K}^\textsf{{rwa}}$ models for $\omega_d/2\pi=5.89$ GHz. Both quantities are divided by their undriven values. Even with the drive frequency close to the transmon resonance, we can identify the clear breakdown of the RWA. The breakdown can be seen even in the dimensionless quantity, the ratio between the Lamb shift and cross-nonlinearity (Fig.~\ref{sm-d}(c)). We attribute the breakdown to the fact that the RWA significantly distorts the undriven Lamb shift and cross-nonlinearity. This results in the distortion of the energy levels of the driven transmon-resonator system.


\subsection{A6. Validity of the approximation.}

We summarize all the approximation used in the derivation of the $\hat{K}$, $\hat{K}_q$, $\hat{K}^\textsf{{rwa}}$, and $\hat{K}_q^\textsf{{rwa}}$ models in this section. Tab.~\ref{table-sm1} and Tab.~\ref{table-sm2} shows the required conditions for the approximation. They also present how good the requirements are satisfied in the experiment. The numbers are approximate values. We assume $\zeta\sim1$ and $\overline{\Omega}_d\sim {\Omega}_d$ to simplify the expression. We also set $\omega_d/2\pi=10$ GHz. The requirements for $\hat{K}$ and $\hat{K}_q$ are satisfied much better than that for $\hat{K}^\textsf{{rwa}}$, and $\hat{K}_q^\textsf{{rwa}}$ in the experiment. For the first requirement in Tab.~\ref{table-sm1}, we set $n=0$ in the subscript of $b^{(\pm)}$. This is because typically the ground and first excited transition frequency is the most important. One can readily identify that the requirements presented in Tab.~\ref{table-sm1} are satistifed much better than those in Tab.~\ref{table-sm2}.

\begin{figure}
    \centering
    \includegraphics[width=0.6\columnwidth]{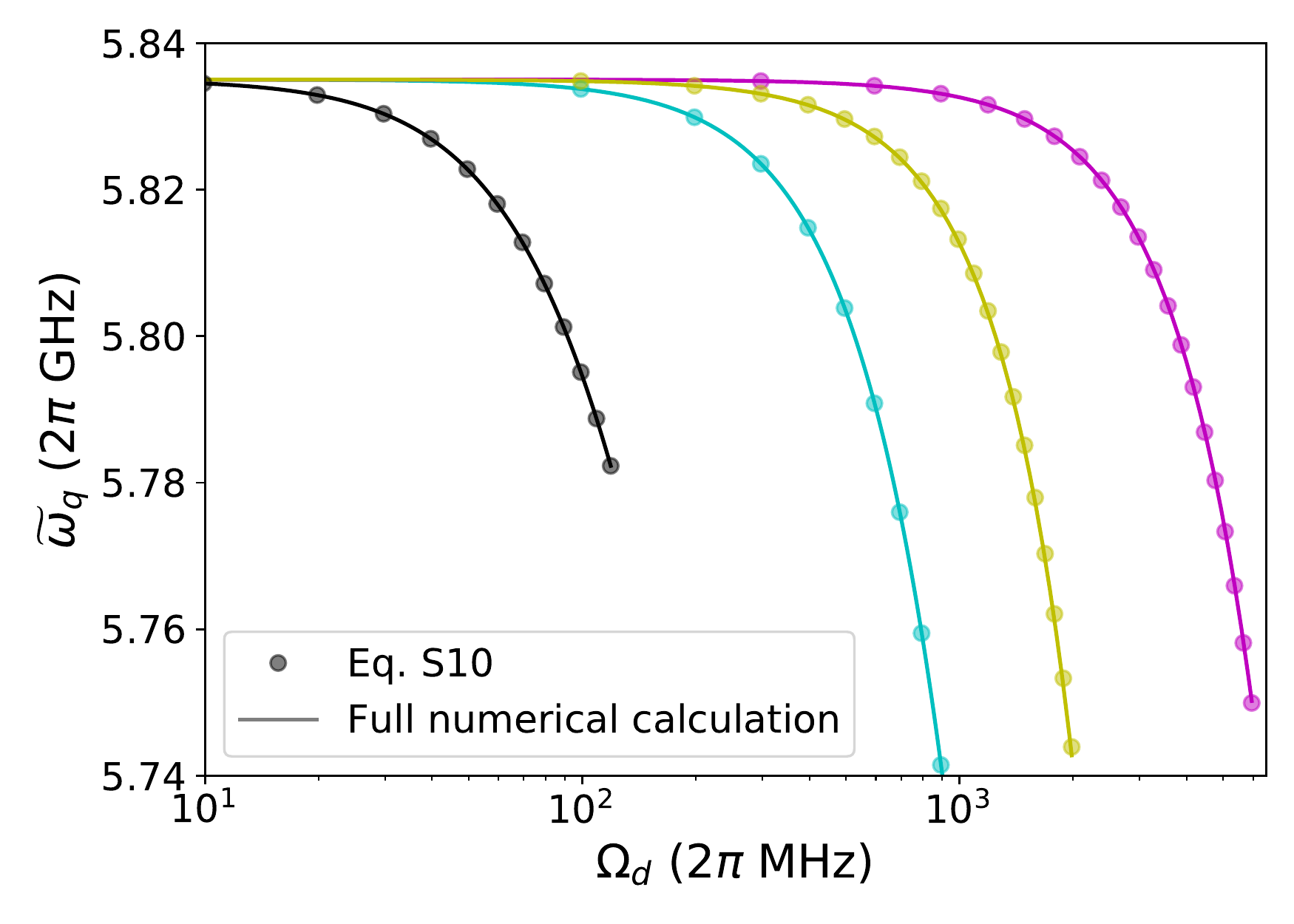}
    \caption{Confirming negligible effect of harmonic generation processes. We plot the renormalized transition frequency from the ground to first eigenstates of the transmon ($\widetilde{\omega}_q$) calculated by two different approaches for various $\omega_d$. Dots refer to the results based on our approximation theory given in Eq.~\ref{eq:10}, which only considers the elements satisfying $|n-m|$ = 1. Lines refer to the results based on full numerical simulation. $\omega_d/2\pi$ for black, cyan, yellow, and magenta is 5.89, 6.5, 3.3, and 10.0 GHz, respectively.}
    \label{sm-e}
\end{figure}

In Fig.~\ref{sm-e}, we also provide the numerical confirmation on our assumption made in Eq.~\ref{eq:6} for analytical calculation of $\hat{K}_q$. We present renormalized transmon resonant frequencies ($\widetilde{\omega}_q$) with respect to drive amplitudes ($\Omega_d$). We consider only the transmon here. Dots are the analytical results based on Eq.~\ref{eq:10}. We only take the components of the interaction terms that meet $|n-m|=1$ into consideration. Lines indicate the results based on numerical calculation without approximation. The detailed approach of the numerical calculation is given in section D.

\begin{center}
\begin{table}
 \begin{tabular}{||c c c||} 
 \hline
 Requirements &  Experiment  & Relevant Eq.\\ [0.5ex] 
 \hline\hline
 $|\frac{\Sigma_{qd}}{\Delta_{qd}}| \frac{b_{0,0+1}^{(\pm)}}{b_{0+1,0}^{(\pm)}} \gg 1$ & 220 & Eq.~\ref{eq:6}\\ 
 \hline
 $ \frac{g^2}{2\overline{\omega}_q|\Delta_{rd}|} \ll 1$ & $8.98 \times 10^{-4}$ & Eq.~\ref{eq:3-3}\\ 
 \hline
 $\frac{2\overline{\omega}_q|\Delta_{rd}\Sigma_{qd}|}{g^2|\Delta_{qd}|} \gg 1 $&  4300  & Eq.~\ref{eq:3-3} \\ 
 \hline
$\frac{g\xi_3}{|\Delta_{rd}|} \ll 1$ & $1.03 \times 10^{-3}$  & Eq.~\ref{eq:3-5}\\ 
 \hline 
\end{tabular}
\caption{\label{table-sm1}Approximation criteria for $\hat{K}_q$ and $\hat{K}$.}
\end{table}
\end{center}

\begin{center}
\begin{table}
 \begin{tabular}{|| c c c||} 
 \hline
 Requirements & Experiment  & Relevant Eq. \\ [0.5ex] 
 \hline\hline
 $|\frac{\Sigma_{qd}}{\Delta_{qd}}| \gg 1$& 3.86 &Eq.~\ref{eq:5-2}, Eq.~\ref{eq:5-3}\\ 
 \hline
 $|\frac{\Sigma_{qr}}{\Delta_{qr}}| \gg 1$ & 6.55 &Eq.~\ref{eq:5-3}\\ 
 \hline
 $\frac{\alpha_4}{\omega_q} \ll 1$ & 0.025 &Eq.~\ref{eq:5-2}, Eq.~\ref{eq:5-3}\\ 
 \hline
\end{tabular}
\caption{\label{table-sm2}Approximation criteria for $\hat{K}_q^\textsf{{rwa}}$ and $\hat{K}^\textsf{{rwa}}$.}
\end{table}
\end{center}

\section{B. Renormalized coherence}
We begin with taking $\hat{U}_q$ on the stochastic Hamiltonian, $\hat{U}_q\hat{H}_{\textbf{st}}\hat{U}_q^\dagger=\lambda_{\parallel}(t)\hat{U}_q\hat{n}\hat{U}_q^\dagger+\lambda_{\perp}(t)\hat{U}_q\hat{d}\hat{U}_q^\dagger$.
$\hat{n}$ and $\lambda(t)$ follow the definitions given in the main text.
$\hat{U}_q\hat{d}\hat{U}_q^\dagger$ is already calculated in Sec.~A2.
As already mentioned, only $\hat{U}_q^{(4)}$ mixes up the eigenbasis of the undressed transmon.
Therefore, we set $\hat{U}_q^{(1)}\hat{U}_q^{(2)}\hat{U}_q^{(3)}\approx 1$ in the derivation of $\hat{U}_q\hat{n}\hat{U}_q^\dagger$.
Just taking $\hat{U}_q^{(4,5)} = \hat{U}_q^{(4)}\hat{U}_q^{(5)}$ on $\hat{n}$ is quite straightforward,
\begin{equation}
\begin{split}
\label{eq:6-1}
    \hat{U}_q^{(4,5)}(t)\hat{n}\hat{U}_q^{(4,5)^\dagger}(t) = \sum_{n,m,k} \ket{m}\langle  {m}\widetilde{\ket{n}}\widetilde{\bra{n}}{k}\rangle\bra{k}e^{-i(m-k)\omega_d t}.
\end{split}
\end{equation}
We note $\widetilde{\ket{n}} = \hat{U}_q^{(4)}\ket{n}$ as a reminder.
With a definition of $\widetilde{n}_{mk} = \sum_{n}\langle{m}\widetilde{\ket{n}}\widetilde{\bra{n}}{k}\rangle$, the transformed $\hat{n}$ operator is expressed
\begin{equation}
\begin{split}
\label{eq:6-2}
    \hat{U}_q(t)\hat{n}\hat{U}_q^\dagger(t) \approx \sum_{n,m} \widetilde{n}_{nm}\ket{n}\bra{m}e^{-i(n-m)\omega_d t}.
\end{split}
\end{equation}
The stochastic Hamiltonian on the renormalized basis is then expressed
\begin{equation}
\begin{split}
\label{eq:6-2}
    \hat{U}_q\hat{H}_{\textbf{st}}\hat{U}_q^\dagger & \approx \\
    & \lambda_{\parallel}(t) \sum_{n,m} \widetilde{n}_{nm}\ket{n}\bra{m}e^{-i(n-m)\omega_d t} + \lambda_{\perp}(t) \sum_{n,m} \widetilde{d}_{nm}^{(-)}\ket{n}\bra{m}e^{i(m-n-1)\omega_d t} - \lambda_{\perp}(t)\sum_{n,m} \widetilde{d}_{nm}^{(+)}\ket{n}\bra{m}e^{i(m-n+1)\omega_d t}.
\end{split}
\end{equation}
Collecting the terms that concern the decay from $\ket{1}$ to $\ket{0}$ yields
\begin{equation}
\begin{split}
\label{eq:6-3}
    \hat{U}_q\hat{H}_{\textbf{st}}\hat{U}_q^\dagger \rightarrow \ket{0}\bra{0}\hat{U}_q\hat{H}_{\textbf{st}}\hat{U}_q^\dagger\ket{1}\bra{1} = \underbrace{\left[ \lambda_{\parallel}(t)\widetilde{n}_{01}e^{i\omega_d t} + \lambda_{\perp}(t)\widetilde{d}_{01}^{(-)} - \lambda_{\perp}(t)\widetilde{d}_{01}^{(+)}e^{2i\omega_d t}\right]}_{=\widetilde{\lambda}_{\perp}(t)} \ket{0}\bra{1} .
\end{split}
\end{equation}
Also, collecting the terms that concern the energy level fluctuations between the $\ket{1}$ and $\ket{0}$ transition yields
\begin{equation}
\begin{split}
\label{eq:6-4}
    \hat{U}_q\hat{H}_{\textbf{st}}\hat{U}_q^\dagger \rightarrow \ket{1}\bra{1}\hat{U}_q\hat{H}_{\textbf{st}}\hat{U}_q^\dagger\ket{1}\bra{1} - \ket{0}\bra{0}\hat{U}_q&\hat{H}_{\textbf{st}}\hat{U}_q^\dagger\ket{0}\bra{0}= \\ \frac{1}{2}\underbrace{\left[ \lambda_{\parallel}(t)(\widetilde{n}_{11}-\widetilde{n}_{00})  -2i\lambda_{\perp}(t)(\widetilde{d}_{11}^{(-)}-\widetilde{d}_{00}^{(-)}) \sin{\omega_dt}\right]}_{=\widetilde{\lambda}_{\parallel}(t)} &(\ket{1}\bra{1} - \ket{0}\bra{0}).
\end{split}
\end{equation}
The renormalized noise spectra is given by
\begin{equation}
\begin{split}
\label{eq:6-5}
    S_{\widetilde{\lambda}_{\perp}}(\omega)=&\frac{1}{2\pi}\int d\tau e^{-i \omega\tau}\left \langle \lambda_{\parallel}^*(t)\lambda_{\parallel}(t+\tau)\widetilde{n}_{01}^2e^{i\omega_d \tau} + \lambda_{\perp}^*(t)\lambda_{\perp}(t+\tau)\widetilde{d}_{01}^{(+)2}e^{2i\omega_d \tau} + \lambda_{\perp}^*(t)\lambda_{\perp}(t+\tau)\widetilde{d}_{01}^{(-)2}\right \rangle
    \\=& \widetilde{n}_{01}^{2}S_{\lambda_{\parallel}}(\omega-\omega_d) + \widetilde{d}_{01}^{(-)2}S_{\lambda_{\perp}}(\omega)+\widetilde{d}_{01}^{(+)2}S_{\lambda_{\perp}}(\omega-2\omega_d).
 \end{split}
\end{equation}
\begin{equation}
\begin{split}
\label{eq:6-5}
    S_{\widetilde{\lambda}_{\parallel}}(\omega)=&\frac{1}{2\pi}\int d\tau e^{-i \omega\tau}\left \langle \lambda_{\parallel}^*(t)\lambda_{\parallel}(t+\tau)(\widetilde{n}_{11}-\widetilde{n}_{00})^2 + \lambda_{\perp}^*(t)\lambda_{\perp}(t+\tau)(\widetilde{d}_{11}^{(-)}-\widetilde{d}_{00}^{(-)})^2(e^{i\omega_d\tau}+e^{-i\omega_d\tau})\right \rangle
    \\=& (\widetilde{n}_{11}-\widetilde{n}_{00})^2S_{\lambda_{\parallel}}(\omega)+(\widetilde{d}_{11}^{~(-)}-\widetilde{d}^{~(-)}_{00})^2S_{{\lambda}_{\perp}}(\omega_d).
 \end{split}
\end{equation}
In the above procedures, we assume that the transverse and longitudinal noise are uncorrelated, that is, $\left \langle \lambda_{\perp}(0)\lambda_{\parallel}(t) \right \rangle =0$.
In the low temperature limit ($\omega_q \gg k_{B}T$), we can ignore absorption process, and thus $S_{\lambda_{\perp}}(\omega)=0$ for $\omega<0$.
With an assumption that the the longitudinal noise $\lambda_{\parallel}(t)$ occurs only at low frequencies, we neglect the $S(\widetilde{\omega}_q^0-\omega_d)\approx 0$ in our cases (cancelled-out).
Finally, we have the relations below
\begin{equation}
\begin{split}
\label{eq:6-6}
   &\frac{1}{\widetilde{T}_1} = \pi \left[ {S}_{\widetilde{\lambda}_{\perp}}(\widetilde{\omega}_q^0) \right] =  \pi \left[S_{{\lambda}_{\perp}}(\widetilde{\omega}_{q}^{0})\widetilde{d}^{~(-) 2}_{01} + \cancelto{\approx0}{S_{{\lambda}_{\parallel }}(\widetilde{\omega}_{q}^{0}-\omega_d)\widetilde{n}_{01}^2} \right], \\
   &\frac{1}{\widetilde{T}_{\varphi }} = \pi \left[ {S}_{\widetilde{\lambda}_{\parallel}}(0) \right] = \pi\left [S_{{\lambda}_{\parallel }}(0)  (\widetilde{n}_{11}-\widetilde{n}_{00})^2+S_{{\lambda}_{\perp}}(\omega_d)(\widetilde{d}_{11}^{~(-)}-\widetilde{d}^{~(-)}_{00})^2\right].
\end{split}
\end{equation}

The calculation so far only considers the transitions within the transmon. In practice, the transmon is coupled to the resonator, and therefore the resonator state should also be considered in the calculation. Then, $\widetilde{d}^{(\pm)}_{nm}$ and $\widetilde{n}^{(\pm)}_{nm}$ in Eq.~\ref{eq:6-6} are replaced with $\widetilde{d}^{~k(\pm)}_{nm}$ and $\widetilde{n}^{~k(\pm)}_{nm}$, which are matrix components of $\hat{U}(t)\hat{d}\hat{U}^\dagger(t)$ and $\hat{U}(t)\hat{n}\hat{U}^\dagger(t)$ when $n_r=k$. 
The same logic can also be applied for $\widetilde{\Omega}_{R}$.
In our case, the resonator is almost empty during the time-domain measurement, and thus, we set $k=0$.

Unless the drives induce resonant sideband transitions between the transmon and resonator, we empirically find that  $\widetilde{d}^{~0(\pm)}_{nm}$ and $\widetilde{n}^{~0(\pm)}_{nm}$ with $\omega_d$ are approximately the same as $\widetilde{d}^{(\pm)}_{nm}$ and $\widetilde{n}^{(\pm)}_{nm}$ with $\omega_d - L_q$. We use this approximation in the calculation presented in the main text. 

Please note that we neglect the effects of the finite resonator linewidth $\kappa$ to the renormalized coherence times of the transmon. In fact, $\kappa$ can also affect the $\widetilde{T}_1$, and $\widetilde{T}_2$, as expected in \cite{multi} using perturbative expansions. Quantitatively describing these effects within our theoretical framework is beyond the scope of this work.

\section{C. Calculated dipole elements}
The eigenbasis renormalization in this paper is mainly investigated through the properties related to the dipole elements $\widetilde{d}^{(\pm)}_{mn}$, which we define in the main text.
These concern many essential properties of the microwave-dressed transmon, such as Lamb shifts, cross-nonlinearities, Rabi frequencies, and energy relaxation times, all of which are essential to precisely estimate the quantum dynamics.
We can expect that the dipole elements behave similarly to those of TS systems as $\Delta_{qd} / \chi_q \rightarrow 0$, and $\Delta_{qd} = \omega_q - \omega_d $.
In the opposite limit ($|\Delta_{qd}| / \chi_q \gg 1$), on the other hand, the  dipole elements will be nearly invariant as increasing the drive amplitudes and this behavior resembles those of linear resonators.
That is to say, the transmon exhibits the two-face-like behavior in both limits.
However, even in the limit of $\Delta_{qd} / \chi_q \rightarrow 0$, one should take the higher energy levels into account, which result in the different behavior of the transmon compared with TS systems.
This is already revealed in Fig.~1 and Fig.~2 in the main text.
\begin{figure}
    \centering
    \includegraphics[width=0.7\columnwidth]{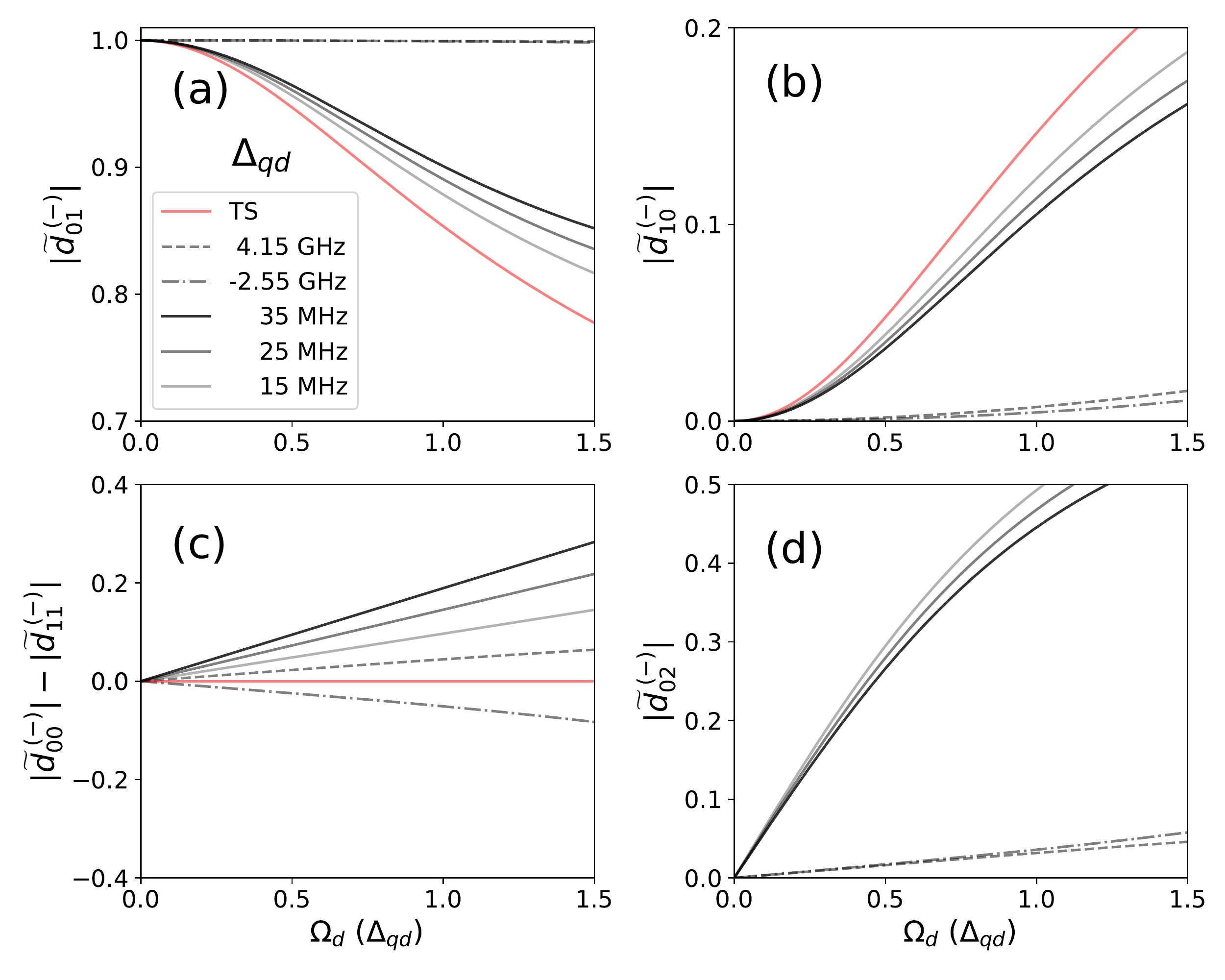}
    \caption{Calculated elements of $\hat{d}^{(-)}$ with respect to the drive amplitudes $\Omega_d$ for various detunings between the transmon and drive fields ($\Delta_{qd}$). $\Omega_d$ is normalized by $\Delta_{qd}$.}
    \label{dipole}
\end{figure}

We confirm this expectation in this section. 
Fig.~\ref{dipole} present the matrix elements of $\widetilde{d}^{(-)}$ on $\widetilde{\ket{n}}$ basis.
Only the elements that have leading effects on the ground and the first excited states of the transmon are presented.
We investigate various transmon–drive field detunings ($\Delta_{qd}$), and also present the case with the TS system model.
For TS systems, $\widetilde{d}$ is hardly dependent on $\Delta_{qd}$ when plotted with respect to $\Omega_d/\Delta_{qd}$.
For transmons, one can readily see the distinct trends depending on $\Delta_{qd}$.
For the small $\Delta_{qd}$, their behavior becomes closer to that of the TS system.
In the opposite limit, however, the change in $\widetilde{d}$ with respect to $\Omega_d$ comes closer to that of linear resonators.
In all cases presented in Fig.~\ref{dipole}, we can readily confirm the expected trends.
One exceptional case is when $\Delta_{qd}/2\pi=-2.55$ GHz (double-dashed lines).
In Fig.~\ref{dipole}(b-d), we can see significant changes in the dipole elements with this drive frequency, although $\Delta_{qd} \gg 1$ holds.


In calculating the dipole elements in Fig.~3 and Fig.~4 of the main text, we consider the renormalization of the transmon's parameters induced by the resonator. This is simply done by replacing the drive frequency $\omega_d$ by $\omega_d + L_{q}$, where $L_{q}$ is the Lamb shift of the undressed transmon. More rigorously, we can also calculate the transformed dipole operator in the transmon–resonator coupled basis. However, this approach requires us to further expand our theoretical derivation from what was presented in Sec.~A, which would take a significant effort. Therefore, we keep this approach from the scope of this paper.

\section{D. Numerical calculation}
\begin{figure}
     \centering
     \includegraphics[width=0.5\columnwidth]{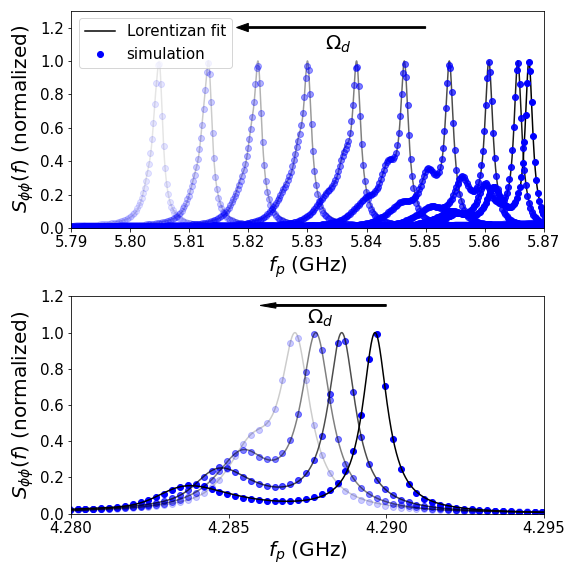}
     \caption{Calculated absorption spectra ${S}_{\phi\phi}(f)$ (circles) of the microwave-dressed transmon (upper) and resonator (lower). Here, $f=\omega/2\pi$.
     The spectra is normalized such that the maximum values become unity. $\omega_d/2\pi$ in the calculation is 5.89 GHz. 
     We set a thermal population in the transmon and resonator of 0.07 to identify cross-nonlinearities between them.
     The other parameters are the same as those in the experiments. The opacity represents the strength of the drive fields. The solid lines are triple Lorentzian fits.}
     \label{fig-sm-spectra}
\end{figure}

Independent of the analytical calculation presented in Sec. A, we numerically calculate the absorption spectra of the microwave-dressed transmon.
The transmon qubit can be modeled as a nonlinear mechanical resonator, replacing the Cooper-pair number ($\hat{N}$) and phase ($\hat{\phi}$) operators by the position ($\hat{x}$) and momentum operators, respectively.
For mechanical systems, the absorption spectra in the linear response regime  is given by Fourier transform of the auto time-correlation function of the position (${S}_{xx}(\omega)$), as defined by Eq.~12 in \cite{cQEMD},
Using this analogy between mechanical and circuit oscillators,
we can deduce that the absorption spectra of the transmon is ${S}_{\phi\phi}(\omega)$, where $\hat{x}$ in Eq.~12 of \cite{cQEMD} is replaced by $\hat{\phi}$, like below,
\begin{equation}
\begin{split}
\label{eq:spec}
  {S}_{\phi\phi}(\omega) = \frac{1}{2\pi}\int d\tau e^{i\omega\tau}\left \langle \hat{\phi}(t+\tau)\hat{\phi}(t) \right \rangle.
\end{split}
\end{equation}
The bracket $\left \langle \dots \right \rangle$ denotes the ensemble average, which can be substituted for the time average for ergodic systems. 
As many time-periodically driven mechanical systems, typical superconducting qubits can also be considered Ergodic. 
The time evolution of $\hat{\phi}(t)$ can be obtained by solving the equation below,
\begin{equation}
\begin{split}
   	\frac{d\hat{\phi}}{dt} =  \vphantom{\sum_n} & -{i} \left[ \hat{H}(t), \hat{\phi}\right]\\
	& + \mathcal{D}[\Gamma^{n,n+1}\sum\ket{n}\bra{n+1}]\hat{\phi} +  \mathcal{D}[\Gamma^{n,n+1}_{\varphi}\sum(\ket{n+1}\bra{n+1})-\ket{n}\bra{n})]\hat{\phi}.
\end{split}
\label{model}
\end{equation}
$\mathcal{D}[\mathcal{\hat{O}}]\hat{\phi}$ is defined by $2\mathcal{\hat{O}}\hat{\phi}\mathcal{\hat{O}}^\dagger - \mathcal{\hat{O}}^\dagger\mathcal{\hat{O}}\hat{\phi}-\hat{\phi}\mathcal{\hat{O}}^\dagger\mathcal{\hat{O}}$.
$\Gamma^{n,n+1}$ and $\Gamma^{n,n+1}_{\varphi}$ are the decay and pure dephasing rates between $\ket{n}$ and $\ket{n+1}$, respectively.
Fig.~\ref{fig-sm-spectra} shows the calculated ${S}_{\phi\phi}(\omega)$ with the same parameters as those in the experiments.
We use the ``correlation-2op-1t" function in QuTip \cite{Qutip1, Qutip2} for numerical calculation.
$\omega_d/2\pi$ in the calculation is set to 5.89 GHz.
We intentionally apply a thermal population (0.07) to the transmon and resonator modes to resolve the cross-nonlinearities.
We can clearly confirm the Stark shifts and renormalization in the cross-nonlinearities from the spectra.
The calculation results based on ${S}_{\phi\phi}(\omega)$ are presented in Fig. 1 in the main text as dashed lines.
We also use ${S}_{\phi\phi}(\omega)$ when calibrating the drive amplitudes in the experiments.
When calculating the time-evolution in the time-correlation function of Eq.~\ref{eq:spec}, we use an unnormalized Lindbladian. 
Also, we set $t=0$ in the correlation function since we find that any $t \in [0,2\pi]$ yields the same transition frequencies.
These choices may lead to incorrect linewidths in the calculated spectra. 
Therefore, we only extract the frequency information from the numerical calculations in this work.

\section{E. Experiment}
\subsection{E1. Device fabrication}
\begin{figure}
     \centering
     \includegraphics[width=0.9\columnwidth]{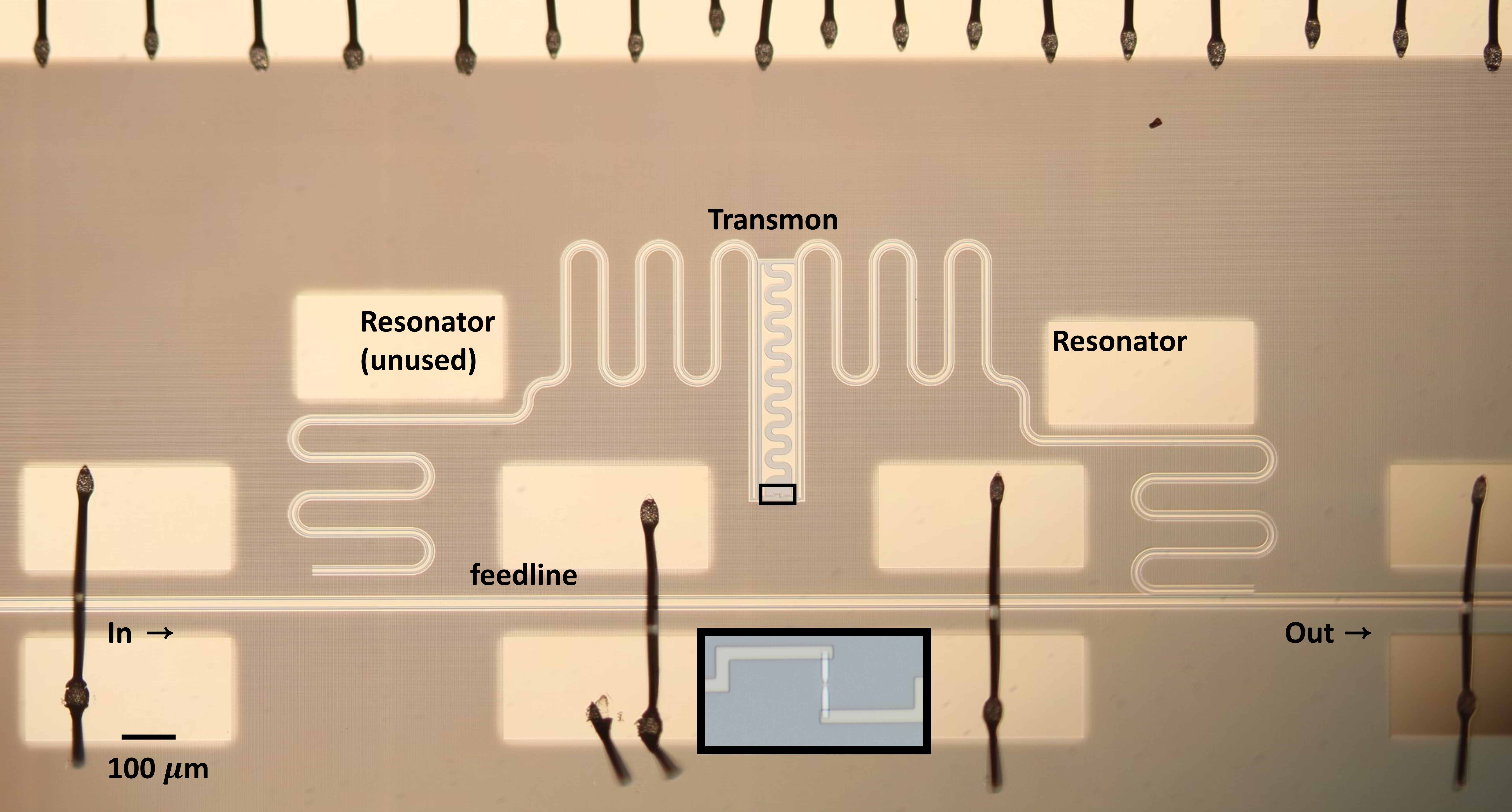}
     \caption{Optical microscopy of the device used in this work. The large rectangle is a magnified view of the small rectangle.}
     \label{fig-sm-device}
\end{figure}

An optical microscopy image of the device is shown in Fig.~\ref{fig-sm-device}.
The transmon is coupled to two coplanar waveguide resonators. The right one is however weakly coupled to the transmon (cross-nonlinearity is less than 100 kHz), and thus we do not take it into account in this paper.
The other ends of the resonators are inductively coupled to the center feedline.
There is no separate drive line for the transmon. Instead, we off-resonantly drive the resonator, which amounts to driving the transmon as long as the itinerant resonator photon number is sufficiently small. See E3 for a detailed proof.
The qubit and resonators were patterned on a 100 nm niobium titanium nitride (NbTiN) film on a silicon substrate \cite{SRON}. 
The Josephson junction of the transmon is an Al-AlOx-Al junction (insets) fabricated using double-angle shadow evaporation.

\subsection{E2. Cryogenic setup}
The device is mounted at the mixing chamber plate of a dilution fridge (Bluefors LD-400) and cooled down to 10mK. We use in total two lines for the experiments, one is for input and the other is for output signals, respectively. To block the thermal noise from the upper plates, we put 10/10/10/20dB attenuators at the 4K/700mK/100mK/MXC plates of the input line, and two isolators at the MXC plate of the output line.
We install a HEMT amplifier (LNF-LNC) at the 4K stage of the output line.
The device is surrounded by radiation and magnetic shields, which are comprised of copper, aluminum, and cryoperm (See \cite{bann} for the detailed design).

\subsection{E3. Device model and parameters}
\label{mode}
This section is allocated to explain how our system can be reduced to the circuit diagram in Fig.1(a) in the main text.
In the experiment, the drive field is actually applied to a resonator through the feedline, and hence the Hamiltonian is then expressed as below, 
\begin{equation}
\begin{split}
\label{eq:bare1}
  \hat{H} = 4E_C\hat{N} - E_J\cos\hat{\phi} + g\hat{d}(\hat{a}-\hat{a}^{\dagger}) + \Omega_d (\hat{a}+\hat{a}^{\dagger}) \cos \omega_d t.
\end{split}
\end{equation}
We apply a displacement operator, $\hat{U}_{dis}(t)=e^{\xi(t)\hat{a}^\dagger-\xi^{*}(t)\hat{a}}$, where
$\xi(t)=\frac{\Omega_{d}}{2\Delta_{rd}}e^{-i\omega_{d}t}+\frac{\Omega_{d}}{2\Sigma_{rd}}e^{i\omega_{d}t}$.
Here, $\Delta_{rd}$ and $\Sigma_{rd}$ are $\omega_r - \omega_d$ and $\omega_r + \omega_d$, respectively.
$\hat{U}_{dis}(t)$ simply displaces the field operator $\hat{a}$($\hat{a}^\dagger$) by -$\xi$(-$\xi^{*}$), while eleminating the time-perodic terms in Eq.~\ref{eq:bare1}.
The transformed Hamiltonian, $\hat{U}_{dis}(t)[\hat{H}-i\partial_t]\hat{U}_{dis}^{\dagger}(t)$ is then expressed
\begin{equation}
\begin{split}
\label{eq:bare2}
  \hat{U}_{dis}(t)[\hat{H}-i\partial_t]\hat{U}_{dis}^{\dagger}(t) = 4E_C\hat{N} - E_J\cos\hat{\phi} + g\hat{d}(\hat{a}-\hat{a}^{\dagger}) - g\hat{d}(\xi(t) - \xi^{*}(t)).
\end{split}
\end{equation}
The time-dependent part of the last term can be rewritten as $\xi(t) - \xi^{*}(t) = i(\frac{\Omega_d}{\Delta_{rd}}-\frac{\Omega_d}{\Sigma_{rd}})\sin\omega_d t$.
By redefining $ig(\frac{\Omega_d}{{\Delta}_{rd}}-\frac{\Omega_d}{\Sigma_{rd}})\rightarrow\Omega_d$, we obtain the drive term $\hat{H}_d$ defined in the main text.

\subsection{E4. Frequency-domain measurement and analysis}
\begin{figure}
     \centering
    \includegraphics[width=0.7\columnwidth]{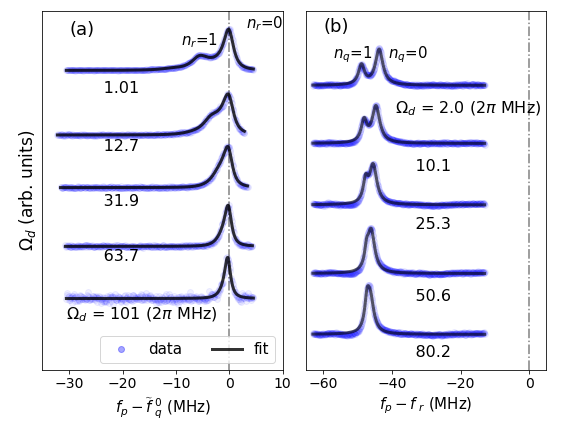}
     \caption{Frequency-domain experiment. Qubit two-tone spectroscopy (a) and resonator transmission data (b) with $\omega_d/2\pi$=5.89 GHz. Blue circles and dark solid lines denote the experimental data and theoretical fits, respectively.}
     \label{fig-sm-fdexp}
\end{figure}

We used a 4-port vector network analyzer (Keysight N5222A) for transmission spectroscopy and signal generator (Keysight N5183B) to feed the microwave drive fields for two-tone spectroscopy. Some of the normalized two-tone spectroscopy and resonator transmission spectra are given in Fig.~\ref{fig-sm-fdexp}.
The drive frequency is 5.89 GHz in the experiment.
One can identify the apparent changes in the spectra with increasing drive amplitudes.
The noise becomes severe as increasing $\Omega_d$ in Fig.~\ref{fig-sm-fdexp}(a). 
This is because the readout fidelity decreases as it becomes hard to distinguish the qubit states in the two-tone spectroscopy with increasing $\Omega_d$ for given $\omega_d$.
The resonator transmission spectra in Fig.~\ref{fig-sm-fdexp}(b) are obtained after correcting the backgrounds.

For two-tone spectroscopy, we first feed a microwave tone at the resonator frequency, and feed a second tone to excite the transmon. The change in the transmon population is translated to the change in the resonator transmission. Thereby, we can probe the transition between the ground and first excited states of the transmon. The resonator is always occupied with some amount of probe photons during the measurement, which can be identified by multiple peaks in the qubit spectrum indicated by $n_r=0$ and $n_r=1$. From theoretical fitting, the estimated photon number is approximately 0.2, substantially less than unity. We can also confirm multiple peaks in the resonator transmission in (b). Each peak corresponds to the different transmon states, $n_q=0$ and $n_q=1$, respectively.

\subsection{E5. Time-domain measurement and analysis}
\begin{figure}
     \centering
     \includegraphics[width=0.5\columnwidth]{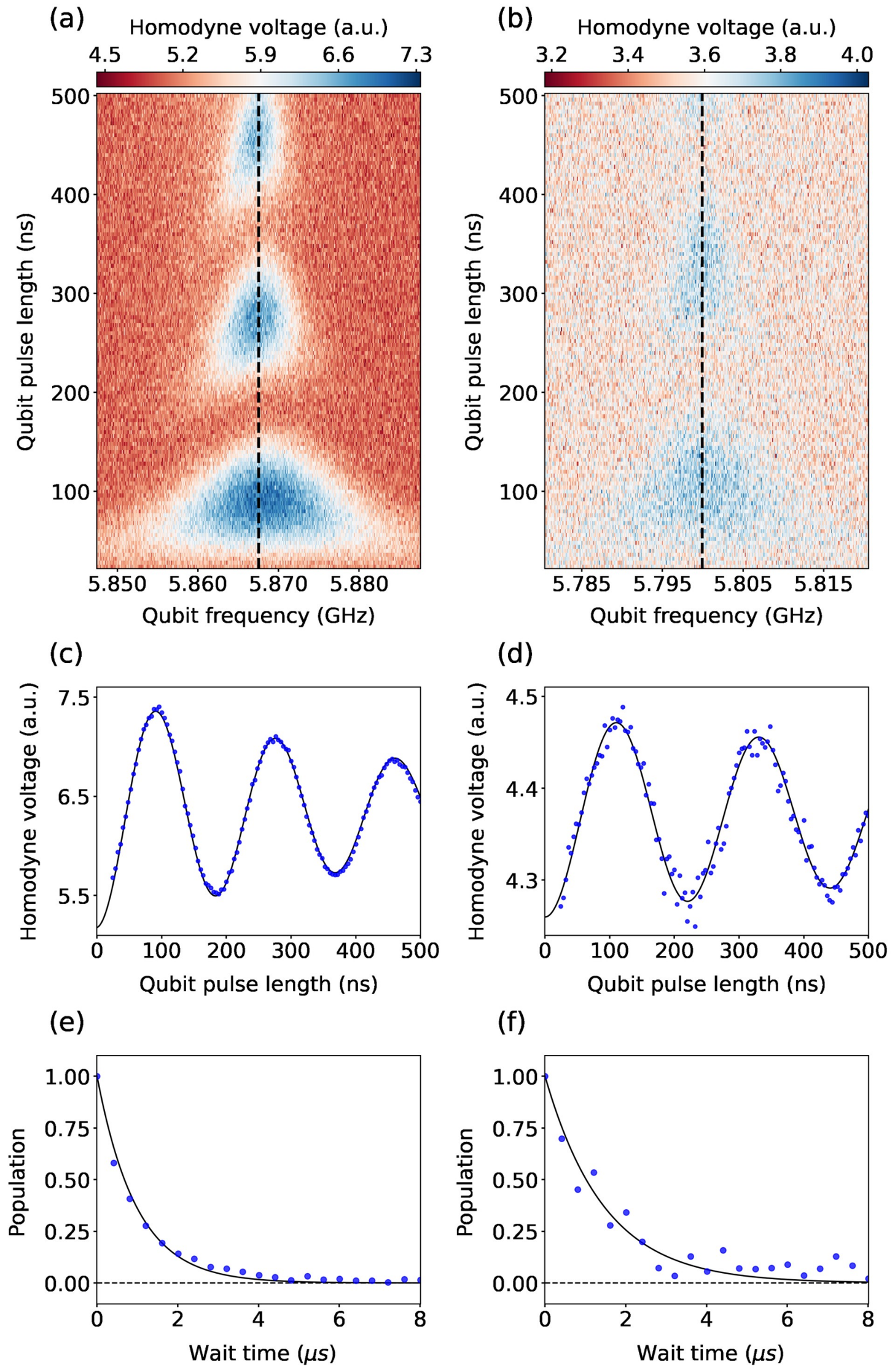}
     \caption{Time-domain experiment. (a-b) Rabi chevrons when $\Omega_d/2\pi$=0 and $\Omega_d/2\pi\approx100$ MHz, respectively. $\omega_d/2\pi$ is set  by 5.89 GHz. The Rabi oscillations and energy relaxations that correspond to the vertical lines are given in (c-d) and (e-f), respectively. When estimating the population in (e-f), we neglect the decay during the qubit excitation.}
     \label{fig-sm-tdexp}
\end{figure}
In time-domain experiments, we use Quantum Machines OPX to generate and capture the pulse signals at IF bandwidth. The qubit control and readout signals are up/down-converted to RF/IF bandwidth by IQ mixers.
We use SGS100A (Rohde-Schwarz) and E8257D (Keysight) to provide local oscillators for the qubit control and readout signal, respectively.
We present the Rabi Chevron data without (a) and with drive field (b) in Fig.~\ref{fig-sm-tdexp}.
The cross sections of the dashed vertical lines are shown in Fig.~\ref{fig-sm-tdexp}(c-d).
The energy relaxation time measurements that correspond to the vertical lines are given in Fig.~\ref{fig-sm-tdexp}(e-f).
The poor contrast in Fig.~\ref{fig-sm-tdexp}(b) and noisy features in Fig.~\ref{fig-sm-tdexp}(d,f) are attributed to the reduced readout fidelity as a result of the decrease in $\chi_{qr}$. 
We can already identify the slight changes in the Rabi oscillation period and energy relaxation time by carefully looking at the fitting curves.

\begin{figure}
     \centering
     \includegraphics[width=0.6\columnwidth]{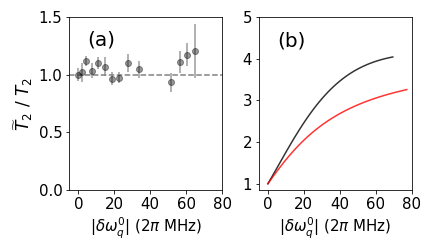}
     \caption{Renormalized coherence time ($\widetilde{T}_{2}$). $\widetilde{T}_2/T_2$ versus $|\delta\omega_q^0|$ for $\omega/2\pi$=10 GHz (a) and 5.89 GHz (b), respectively. 
     Only the theoretical estimations are given in (b) due to the experimental limitation. See the corresponding text for the detailed reason.}
     \label{fig-sm-T2}
\end{figure}

\section{F. $T_2$ measurement}
In Fig.~\ref{fig-sm-T2}, we also investigate the renormalized coherence time $\widetilde{T}_{2}$.
In our device, $T_2\approx0.5\mu s$, and $T_1\approx1\mu s$. Therefore, $\widetilde{T}_{2}$ is mainly determined by $\widetilde{T}_{\phi}$.
In Fig.~\ref{fig-sm-T2}(a), $\widetilde{T}_{2}$ seems almost constant with respect to $|\delta\omega_q^0|$.
We only present the theoretical estimation in Fig.~\ref{fig-sm-T2}(b). With $\omega_d/2\pi=5.89$ GHz, Ramsey experiments require a lot of time in our case when increasing the drive amplitudes. 
This is because the readout fidelity drops significantly as $\chi_{qr}$ becomes smaller, which can already be identified in Fig.~1 and Fig.~2 of the main text.
For this reason, we present only the theoretical expectations in Fig.~\ref{fig-sm-T2}(b).

\section{G. Extended calculations for renormalized quantities}
\begin{figure}
     \centering
     \includegraphics[width=0.8\columnwidth]{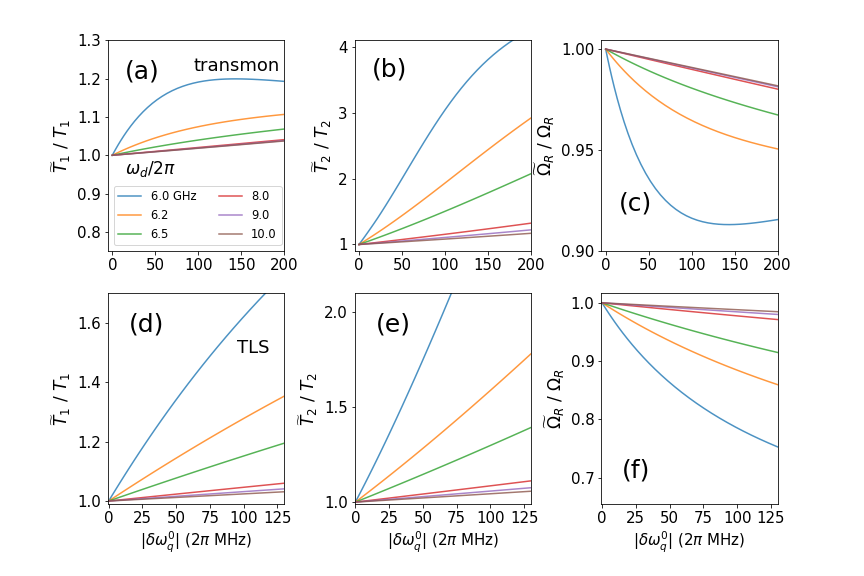}
     \caption{Calculated $\widetilde{T}_{1}$, $\widetilde{T}_{2}$, and $\widetilde{\Omega}_{R}$ for various $\omega_d$. Transmon (a-c) and two-level system case (d-f) are separately presented.}
     \label{fig-sm-T_fscan}
\end{figure}

In the main text, we present the experimental data of $\widetilde{\Omega}_{R}$ and $\widetilde{T}_1$ for certain values of $\omega_d$. This is because of the formidable time cost as the readout fidelity decreases (See the explanation in Sec.~E4). For some unknown technical reasons, time-domain measurement requires longer times to obtain sufficient signal-to-noise ratio than the frequency-domain measurement does in our case. In Fig.~\ref{fig-sm-T_fscan}, we present the extended calculation results of the renormalized $\widetilde{T}_1$,  $\widetilde{T}_2$, and $\widetilde{\Omega}_{R}$ for various $\omega_d$ to help readers see the expected tendency.

\end{document}